\documentclass[pre,floatfix,longbibliography,superscriptaddress,twocolumn,final,notitlepage]{revtex4-2}

\RequirePackage{amsmath}
\RequirePackage{hyperref} 
\RequirePackage{amssymb}
\usepackage{graphicx}
\usepackage{chngcntr}

\usepackage[nameinlink,capitalize]{cleveref}

\makeatletter \newcommand\setcurrentname[1]{\def\@currentlabelname{#1}} 

\newcommand{\mtcov}{\mbox{\small MTCOV}}
\newcommand{\model}{\mbox{\small PIHAM}}

\newcommand{\A}{\boldsymbol{A}}
\newcommand{\X}{\boldsymbol{X}}
\newcommand{\T}{\boldsymbol{\Theta}}
\newcommand{\p}{\boldsymbol{\theta}}
\newcommand{\U}{\boldsymbol{U}}
\newcommand{\V}{\boldsymbol{V}}
\newcommand{\w}{\boldsymbol{W}}
\newcommand{\Wone}{\boldsymbol{W^1}}
\newcommand{\Wtwo}{\boldsymbol{W^2}}
\newcommand{\ui}{\boldsymbol{U_i}}
\newcommand{\vi}{\boldsymbol{V_i}}
\newcommand{\vj}{\boldsymbol{V_j}}
\newcommand{\wl}{\boldsymbol{W^{\ell}}}
\newcommand{\h}{\boldsymbol{H}}
\newcommand{\hx}{\boldsymbol{H_{\cdot x}}}
\newcommand{\Aone}{\boldsymbol{A^1}}
\newcommand{\Atwo}{\boldsymbol{A^2}}
\newcommand{\Athree}{\boldsymbol{A^3}}
\newcommand{\Abin}{\boldsymbol{A^{\ell}}}
\newcommand{\Adist}{\boldsymbol{A^7}}
\newcommand{\xcaste}{\boldsymbol{X_{\cdot 1}}}
\newcommand{\xrel}{\boldsymbol{X_{\cdot 2}}}
\newcommand{\xedu}{\boldsymbol{X_{\cdot 3}}}
\newcommand{\id}{\boldsymbol{i}}
\newcommand{\m}{\boldsymbol{\mu}}
\newcommand{\s}{\boldsymbol{\Sigma}}
\newcommand{\al}{\boldsymbol{\alpha}}
\newcommand{\N}{\mathcal{N}}
\DeclareMathOperator{\softmax}{softmax}
\DeclareMathOperator{\logistic}{logistic}
\DeclareMathOperator{\Bern}{Bern}
\DeclareMathOperator{\Pois}{Pois}
\DeclareMathOperator{\Cat}{Cat}
\DeclareMathOperator{\Dir}{Dir}
\DeclareMathOperator{\Lognormal}{Lognormal}
\DeclareMathOperator{\Logitnormal}{Logitnormal}

\newcommand{\avg}{\langle{k}\rangle}

\usepackage{adjustbox}
\usepackage{booktabs}
\renewcommand{\thetable}{\arabic{table}}

\makeatletter
    \renewcommand\@make@capt@title[2]{%
     \@ifx@empty\float@link{\@firstofone}{\expandafter\href\expandafter{\float@link}}%
      {\textbf{#1}}\@caption@fignum@sep#2\quad}%
    \makeatother
   
\makeatletter 
\renewcommand{\fnum@figure}{\textbf{Fig.~\thefigure}}
\renewcommand{\fnum@table}{\textbf{Table~\thetable}}
\makeatother

\begin{document}

\title{Flexible inference in heterogeneous and attributed multilayer networks}

\author{Martina Contisciani}
	\email{martina.contisciani@tuebingen.mpg.de}
	\thanks{Contributed equally.}
	\affiliation{Max Planck Institute for Intelligent Systems, Tübingen 72076, Germany}
\author{Marius Hobbhahn}
	\thanks{Contributed equally.}
	\affiliation{Tübingen AI Center, University of Tübingen, Tübingen 72076, Germany}
\author{Eleanor A. Power}
	\affiliation{Department of Methodology, London School of Economics and Political Sciences, London WC2A 2AE, United Kingdom}
	\affiliation{Santa Fe Institute, 1399 Hyde Park Road, Santa Fe, New Mexico 87501, USA}
\author{Philipp Hennig}
	\affiliation{Tübingen AI Center, University of Tübingen, Tübingen 72076, Germany}
\author{Caterina De Bacco}
	\email{caterina.debacco@tuebingen.mpg.de}
	\affiliation{Max Planck Institute for Intelligent Systems, Tübingen 72076, Germany}

\begin{abstract} 
Networked datasets can be enriched by different types of information about individual nodes or edges. However, most existing methods for analyzing such datasets struggle to handle the complexity of heterogeneous data, often requiring substantial model-specific analysis. In this paper, we develop a probabilistic generative model to perform inference in multilayer networks with arbitrary types of information. Our approach employs a Bayesian framework combined with the Laplace matching technique to ease interpretation of inferred parameters. Furthermore, the algorithmic implementation relies on automatic differentiation, avoiding the need for explicit derivations. This makes our model scalable and flexible to adapt to any combination of input data. We demonstrate the effectiveness of our method in detecting overlapping community structures and performing various prediction tasks on heterogeneous multilayer data, where nodes and edges have different types of attributes. Additionally, we showcase its ability to unveil a variety of patterns in a social support network among villagers in rural India by effectively utilizing all input information in a meaningful way. 
\end{abstract}

\maketitle

\section*{Introduction} 
Networks effectively represent real-world data from various fields, including social, biological, and informational systems. In this framework, nodes within the network correspond to individual components of the system, and their interactions are illustrated through network edges~\cite{newman2018networks}. With the advancement of data collection and representation techniques, networks have evolved to become more versatile and informative. Notably, attributed multilayer networks have emerged as a significant development, allowing the inclusion of additional information related to nodes and edges. This enriches the representation of real-world systems, where nodes naturally have specific characteristics and are connected through different types of interactions. For instance, in social networks, individuals can be described by attributes like age, gender, and height, while engaging in various types of relationships like friendship, co-working, and kinship. 

It is important to treat these systems as multilayer networks, rather than reducing them to a single layer (e.g., through aggregation), to avoid losing valuable information. This is particularly significant when the properties of the entire system cannot be derived from a simple linear combination of the properties of each layer in isolation, or through other forms of aggregation into a single layer representation~\cite{de2023more}. Relevant examples of this non-trivial behavior are observed in interbank markets, patent citation networks, time-dependent networks and brain networks ~\cite{bargigli2015multiplex,higham2022multilayer,mucha2010community,battiston2017multilayer}. A similar rationale applies to datasets with multiple types of edges and node attributes, as explored in this work. Approaches to multilayer networks that consider only one type of attribute -- or none at all -- risk losing critical information.

The analysis of attributed multilayer networks has primarily been tackled using techniques like matrix factorization~\cite{liu2020community, xu2023covariate}, network embedding~\cite{pei2018attributed, cen2019representation}, and deep learning~\cite{cao2018incorporating, park2020unsupervised, han2022node, martirano2022comlhan}. 
While effective for learning low-dimensional node representations for inference tasks, these methods have significant limitations when applied to network data. First, the resulting representations lack inherent interpretability. Since their main focus is prediction tasks, these models do not impose constraints on the parameters to improve interpetability, such as non-negativity or shared parameters across layers. Instead, they often require arbritarily post-processing techniques (e.g., k-means clustering) for interpretation. Second, these methods prioritize node-level tasks (e.g., classification or clustering) and treat edge-level tasks, like link prediction, as secondary, relying on ad hoc mappings from nodes to edges. Lastly, they are typically parameter-heavy and require a substantial amount of training data, making them inefficient and overparametrized for real-world networks, which are typically sparse, label-scarce for~supervised settings, and often provide only a single observed sample.

To overcome these limitations, we adopt a different approach based on probabilistic generative models~\cite{goldenberg2010survey}. Unlike the aforementioned methods, these models provide a principled and flexible framework that incorporates prior knowledge and specific assumptions, resulting in more interpretable representations. Importantly, they explicitly capture the dependencies between nodes and edges, rather than relying on ad hoc mappings. They also account for the inherent uncertainty present in real-world network data~\cite{peel2022statistical}, providing a more robust and comprehensive understanding of this type of data. Furthermore, these models can be applied to perform various network tasks, such as edge and attribute prediction, detecting statistically meaningful network structures, and generating synthetic data.

Our goal is to develop a probabilistic generative model that can flexibly adapt to any attributed multilayer network, regardless of the type of information encoded in the data. Acting as a ``black box'', our method can enable practitioners to automatically analyze various datasets, without the need to deal with mathematical details or new derivations. This approach aligns with some practices in the machine learning community, where black box methodologies have been introduced to simplify the inference of latent variables in arbitrary models~\cite{ranganath2014black, tran2016variational}. In this context, more specific probabilistic methods have been developed to address the challenge of performing inference on heterogeneous data~\cite{valera2020general, nazabal2020handling}. However, these techniques are tailored for tabular data and do not provide a general solution to adapt them to network~data.

Probabilistic generative models specifically designed for attributed networks aim to combine node attributes effectively with network interactions. Existing methods~\cite{tallberg2004bayesian, yang2013community, hric2016network, newman2016structure, white2016mixed, stanley2019stochastic, fajardo2022node, contisciani2020community} have highlighted the importance of incorporating extra information to enhance network inference, resulting in improved prediction performance and deeper insights on the interplay between edge structure and node metadata. However, these models mainly focus on single-layer networks, assume the same generative process for all interactions, and consider only one type of attribute -- typically categorical. These limitations restrict their capability to represent complex scenarios characterized by heterogeneous information. As a consequence, addressing the challenge of effectively incorporating various sources of information and evaluating their collective impact on downstream network inference tasks remains an open issue.

We address this gap by introducing \model, a generative model explicitly designed to perform Probabilistic Inference in directed and undirected Heterogeneous and Attributed Multilayer networks. Our approach differs from previous studies in that \model\ flexibly adapts to any combination of input data, while standard probabilistic methods rely on model-specific analytic derivations that highly depend on the data types given in input. This can dramatically hinder the flexibility of a model, as any small change in the data, e.g., adding a new node attribute or a new type of interaction, usually requires new derivations. As a result, the vast majority of these models work only with one type of edge weight for all layers, and one type of attribute. In contrast, \model\ takes in input any number of layers and attributes, regardless of their data types.

At its core, \model\ assumes the existence of a mixed-membership community structure that drives the generation of both interactions and node attributes. In addition, the inference of the parameters is performed within a Bayesian framework, where both prior and posterior distributions are modeled with Gaussian distributions. Importantly, \model\ employs the Laplace matching technique~\cite{hobbhahn2021laplace} and conveniently maps the posterior distributions to various desired domains, to ease interpretation. For instance, to provide a probabilistic interpretation of the inferred communities, our method properly maps the parameters of a Gaussian distribution into those of a Dirichlet distribution. The latter operates within a positive domain and enforces normalization on a simplex, making it a valuable tool for this purpose. Notably, the inference process is flexible and scalable, relying on automatic differentiation and avoiding the need for explicit derivations. As a result, \model\ can be considered a ``black box'' method, as practitioners only need to select the desired probabilistic model and a set of variable transformation functions, while the remaining calculations and inference are performed automatically. This versatility enables our model to be flexibly applied to new modeling scenarios.

We apply our method on a diverse range of synthetic and real-world data, showcasing how \model\ effectively leverages the heterogeneous information contained in the data to enhance prediction performance and provide richer interpretations of the inferred results.

\section*{Methods}
We introduce \model, a versatile and scalable probabilistic generative model designed to perform inference in attributed multilayer networks. Our method flexibly adapts to any combination of input data, regardless of their data types. For simplicity, in what follows, we present examples with Bernoulli, Poisson, Gaussian, and categorical distributions, which collectively cover the majority of real-world examples. Nevertheless, our model can be easily extended to include new distributions, as well as applied to single-layer networks with or without attributes.

\begin{figure*}[ht!]
\centering
\includegraphics[width=0.9\textwidth]{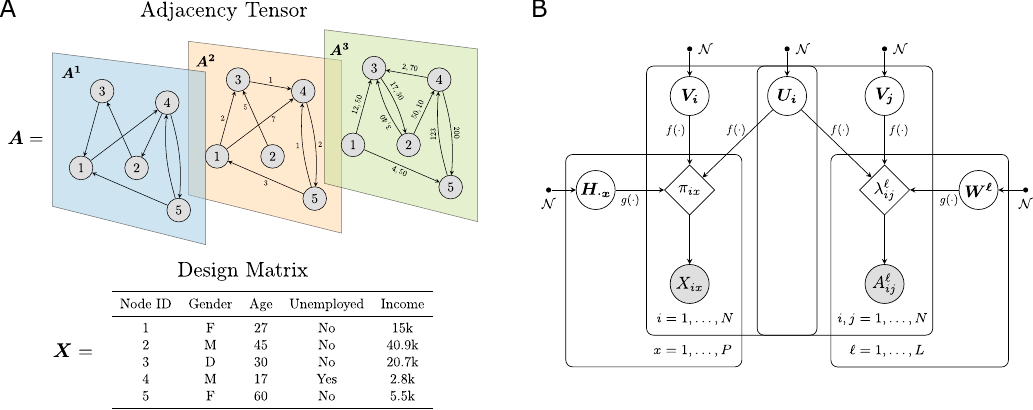}
\caption{Input data and graphical model representation. (A) The attributed multilayer network is represented by the interactions $A_{ij}^{\ell}$ and the node attributes $X_{ix}$. (B) \model\ describes the observed data through a set of latent variables $\T = (\U, \V, \w, \h)$. $\ui$ and $\vi$ respectively depict the communities of node $i$ determined by the out-going and in-coming edges; $\wl$ is the affinity matrix associated to the layer $\ell$ and characterizes the edge density between different community pairs in the given layer; $\hx$ is a $K$-dimensional vector that explains how an attribute $x$ is distributed among the $K$ communities. All latent variables are independent and normally distributed, and $f(\cdot)$ and $g(\cdot)$ are transformation functions to ensure that the expected values $\lambda_{ij}^{\ell}$ and $\pi_{ix}$ belong to the correct parameter space for the various distribution types.}
\label{fig:graphical_model}
\end{figure*}

\subsection*{General framework} 
Attributed multilayer networks provide an efficient representation of complex systems in which the individual components have diverse attributes (often referred to as covariates or metadata) and are involved in multiple forms of interactions. 
Mathematically, these interactions are depicted by an adjacency tensor $\A$ of dimension $L \times N \times N$, where $N$ is the number of nodes common across all $L$ layers. Each entry $A_{ij}^{\ell}$ in this tensor denotes the weight of a directed interaction of type $\ell$ from node $i$ to node $j$. Notably, different layers can incorporate interactions of diverse data types, depending on the nature of the underlying relationship. For instance, in social systems, one layer might represent binary relationships like friendships, another could describe nonnegative discrete interactions such as call counts, and a third might contain continuous real-valued measurements such as geographical distances between locations. In this scenario, the adjacency tensor would be represented as $\A = \{\Aone \in \{0, 1\}^{N \times N}, \Atwo \in \mathbb{N}_0^{N \times N}, \Athree \in \mathbb{R}_+^{N \times N} \}$. Node metadata describes additional information about the nodes. They are stored in a design matrix $\X$ with dimensions $N \times P$, where $P$ is the total number of attributes and the entries $X_{ix}$ represent the value of an attribute $x$ for a node $i$. Similar to network interactions, different attributes can have different data types. An example of input data is given in \cref{fig:graphical_model}A.

\model\ describes the structure of attributed multilayer networks, represented by $\A$ and $\X$, through a set of latent variables $\T$. The goal is to infer $\T$ from the input data. In particular, we want to estimate posterior distributions, as done in a probabilistic framework. These can be approximated as:
\begin{align}\label{eq:general_framework}
    P(\T \,|\, \A, \X) &\propto P(\A, \X \,|\, \T) \, P(\T) \nonumber \\
    & = P(\A \,|\, \T) \, P(\X \,|\, \T) \, P(\T) \, .
\end{align}
In this general setting, the proportionality is due to the omission of an intractable normalization term that does not depend on the parameters. The term $P(\A, \X \,|\, \T) = P(\A \,|\, \T)\, P(\X \,|\, \T)$ represents the likelihood of the data, where we assume that $\A$ and $\X$ are conditionally independent given the parameters. This assumption allows to model separately the network structure and the node metadata. The term $P(\T)$ denotes the prior distributions of the latent variables, which we assume to be independent and Gaussian distributed, resulting in $P(\T) = \prod_{\p \in \T} \N(\p; \m^{\p}, \s^{\p})$. Importantly, we also  make the assumption that the posterior distributions of the parameters can be approximated with Gaussian distributions, for which we have to estimate mean and covariance matrices: $P(\T \,|\, \A, \X) \approx \prod_{\p \in \T} \N(\p; \hat{\m}^{\p}, \hat{\s}^{\p})$.  

In the following subsections, we provide additional details on the role of the latent variables in shaping both interactions and node attributes, as well as the methods for inferring their posterior distributions.

\subsection*{Modelling the network structure} 
The interactions encoded in the adjacency tensor $\A$ are assumed to be conditionally independent given the latent variables, resulting in a decomposition of the likelihood across individual entries $A_{ij}^{\ell}$. This factorization can be further unpacked by explicitly considering the distributions that describe each layer. For instance, in the scenario with binary, count-based, and continuous interactions, we can express the likelihood as follows:
\begin{align}\label{eq:likelihood_network}
    P(\A \,|\, \T) &= \prod_{\ell, i, j} P(A_{ij}^{\ell} \,|\, \T) \nonumber \\
    &= \prod_{\ell \in L_B, i, j} \Bern(A^{\ell}_{ij}; \lambda^{\ell}_{ij}(\T)) \nonumber \\ 
    &\quad \times \prod_{\ell \in L_P, i, j} \Pois(A^{\ell}_{ij}; \lambda^{\ell}_{ij}(\T)) \nonumber \\ 
    &\quad \times \prod_{\ell \in L_G, i, j} \N(A^{\ell}_{ij}; \lambda^{\ell}_{ij}(\T), \sigma^2) \, ,
\end{align}
where $\sigma^2$ is a hyperparameter and $L_B, L_P$, and $L_G$ are the sets of Bernoulli, Poisson, and Gaussian layers, respectively. We assume that each distribution is fully parametrized through the latent variables $\T$ and these explicitly define the expected values $\lambda_{ij}^{\ell}$, regardless of the data type. 

Specifically, we adopt a multilayer mixed-membership model~\cite{debacco2017community}, and describe the observed interactions through $K$ overlapping communities shared across all layers. Following this approach, the expected value of each interaction of type $\ell$ from node $i$ to $j$ can be approximated as:
\begin{equation} \label{eq:lambda}
\lambda_{ij}^{\ell}(\T) \approx \sum_{k, q=1}^K U_{ik} W_{kq}^{\ell} V_{jq} \, ,
\end{equation}
where the latent variables $U_{ik}$ and $V_{jq}$ denote the entries of $K$-dimensional vectors $\ui$ and $\vi$, which respectively represent the communities of node~$i$ determined by the out-going and in-coming edges. In undirected networks, we set $\U = \V$. Moreover, each layer $\ell$ is associated with an affinity matrix $\wl$ of dimension $K \times K$, which characterizes the edge density between different community pairs in the given layer $\ell$. This setup allows having diverse structural patterns in each layer, including arbitrarily mixtures of assortative, disassortative and core-periphery structures.

As a final remark, the approximation in \cref{eq:lambda} arises from a discrepancy between the parameter space of the latent variables and that of the expected values of the distributions. In fact, while all variables are normally distributed, $\lambda_{ij}^{\ell}$ has to satisfy different constraints according to the distribution type. For instance, $\lambda_{ij}^{\ell} \in [0,1] \, \forall \, \ell \in L_B$ and $\lambda_{ij}^{\ell} \in (0,\infty) \, \forall \, \ell \in L_P$. For further details, we refer to the section \nameref{subsec:parameter_space}.

\subsection*{Modelling the node metadata}
Similarly to the network edges, the node metadata are also considered to be conditionally independent given the latent variables. Therefore, when dealing with data that encompass categorical, count-based, and continuous attributes, the likelihood can be formulated as follows:
\begin{align}\label{eq:likelihood_metadata}
    P(\X \,|\, \T) &= \prod_{i, x} P(X_{ix} \,|\, \T) \nonumber \\
    &= \prod_{i, x \in C_C} \Cat(X_{ix}; \pi_{ix}(\T)) \nonumber \\ 
    &\quad \times \prod_{i, x \in C_P} \Pois(X_{ix}; \pi_{ix}(\T)) \nonumber \\ 
    &\quad \times \prod_{i, x \in C_G} \N(X_{ix}; \pi_{ix}(\T), \sigma^2) \, ,
\end{align}
where $C_C, C_P$, and $C_G$ are the sets of categorical, Poisson, and Gaussian attributes, respectively. 

Following previous work~\cite{yang2013community, contisciani2020community, fajardo2022node}, we assume that the attributes are also generated from the node community memberships, thereby creating dependencies between node metadata and network interactions. In particular, we approximate the expected value of an attribute $x$ for node~$i$~as:
\begin{equation} \label{eq:pi}
\pi_{ix}(\T) \approx \frac{1}{2} \sum_{k=1}^K (U_{ik} + V_{ik}) \, H_{kx}  \, ,
\end{equation}
where $\h$ is a $K \times P$-dimensional community-covariate matrix, explaining how an attribute $x$ is distributed among the $K$ communities. For instance, if we consider income as node metadata and expect communities to group nodes with similar income values, then the column vector $\hx$ describes how income varies across groups. It is important to observe that when the attribute $x$ is categorical, the expression in \cref{eq:pi} becomes more complex because it must consider the total number of attribute categories $Z$. We provide additional details in \cref{sec:cat_attribute} of the Supporting Information.

Notice that like $\lambda_{ij}^{\ell}$, $\pi_{ix}$ also needs to satisfy specific constraints depending on the distribution type. We clarify this in the next subsection. 

\subsection*{Parameter space and transformations}\setcurrentname{Parameter space and transformations}\label{subsec:parameter_space}
A key technical aspect of \model\ is the use of Gaussian distributions to model priors and posteriors of the latent variables $\T=(\U, \V, \w, \h)$. This choice simplifies the inference by an additional step that ensures the expected values $\lambda_{ij}^{\ell}$ and $\pi_{ix}$ belong to the correct parameter space for the various distribution types. To achieve this, we apply specific transformation functions to the latent variables, and model the expected values as follows:
\begin{align}
\lambda_{ij}^{\ell}(\T) &= f(\ui) \, g(\wl) \, f(\vj) \label{eq:transformation_lambda} \\
\pi_{ix}(\T) &= \frac{1}{2} \big(f(\ui) + f(\vi)\big) \, g(\hx) \label{eq:transformation_pi} \, .
\end{align} 
The functions $f(\cdot)$ and $g(\cdot)$ can take various forms, as long as they adhere to the required constraints. In our implementation, we select $f(\cdot)$ to be the $\softmax$ function, which is applied to every row of the community membership matrices. This allows interpretability of the communities, as they result in quantities that are positive and normalized to one, as discussed in the section \nameref{subsec:parameter_interpretation}. Meanwhile, the choice of $g(\cdot)$ varies depending on the distribution type, as illustrated in \cref{tab:g_transformations}.

\begin{table}[t!]
\centering
\renewcommand{\arraystretch}{1.1}
\begin{tabular}{lcc}
Distribution & Parameter space & Transformation function \\
\midrule
Bernoulli & $[0, 1]$ & Logistic \\
Poisson & $(0, \infty)$ & Exponential \\
Gaussian & $\mathbb{R}$ & Identity \\
Categorical & $p_z \geq 0 \, \forall z, \sum_{z} p_z= 1$ & Softmax \\
\bottomrule
\end{tabular}
\caption{\label{tab:g_transformations} Functions $g(\cdot)$ used in our implementation to transform the latent variables as defined in \cref{eq:transformation_lambda} and \cref{eq:transformation_pi}. For Gaussian distributions, we model only the mean; for categorical distributions, we apply the $\softmax$ by row, i.e., across categories.}
\end{table}

One might argue that it would be simpler to employ a single link function for $\lambda_{ij}^{\ell}$ and $\pi_{ix}$, rather than applying individually transformations to the latent variables, as done in standard statistical approaches~\cite{mccullagh2019generalized}. However, this may not ensure interpretability of the communities, as we do with the $\softmax$ $f(\cdot)$. In addition, empirically we discovered that the approach outlined in \cref{eq:transformation_lambda} and \cref{eq:transformation_pi} gives more stable results, and it does not result in over- or under-flow numerical errors. Alternatively, another approach considers treating the transformed parameters as random variables and applies the probability transformation rule to compute their posterior distributions~\cite{kucukelbir2017automatic}. While this method is theoretically well-founded, it comes with constraints regarding the choice of the transformation functions, which directly affects the feasibility of the inference process. Conversely, \model\ offers the flexibility to use any set of transformation functions that respects the parameter space of the distribution types given by the network and covariates. 

\subsection*{Posterior inference}
\model\ aims at estimating the posterior distributions of the latent variables, as outlined in \cref{eq:general_framework}, where the normalization term is omitted as it does not depend on the parameters. More precisely, this equation can be reformulated as:
\begin{align}\label{eq:posterior}
    P(\U, \V, \w, \h \, | \, \A, \X) &= P(\A \,|\, \U, \V, \w) \nonumber \\ 
    &\quad \times P(\X \,|\, \U, \V, \h) \nonumber \\ 
    &\quad \times P(\U) \, P(\V) \, P(\w) \, P(\h) \, .
\end{align}
In general, this posterior distribution lacks a closed-form analytical solution and requires the use of approximations.

Common methods for inference in attributed networks typically rely on Expectation-Maximization~(EM)~\cite{dempster1977maximum} or Variational Inference~(VI)~\cite{blei2017variational} techniques. However, these approaches have limitations, as they require model-specific analytic computations for each new term added to the likelihood. For instance, an EM-based approach involves taking derivatives with respect to a given latent variable and setting them to zero. In a specific class of models where the likelihood and prior distributions are compatible, solving the resulting equation for the variable of interest can yield closed-form updates. Nonetheless, for generic models, there is no guarantee of a closed-form solution. Even when this does exist, slight variations in the input data may require entirely new derivations and updates. Consequently, most of these models are designed to handle only a single type of edge weight and a single type of attribute.

In contrast, our model takes a different approach and flexibly adapts to any combination of input data, regardless of their data types. We begin by assuming that the latent variables are conditionally independent given the data, allowing us to model each posterior distribution separately:
\begin{align}\label{}
    P(\U, \V, \w, \h \, | \, \A, \X) &= P(\U \, | \, \A, \X) \, P(\V \, | \, \A, \X) \, \nonumber \\
    &\, \, \, \times P(\w \, | \, \A, \X) \, P(\h \, | \, \A, \X) \, .
\end{align}
Subsequently, we employ a Laplace Approximation~(LA) to approximate each posterior with a Gaussian distribution, resulting in:
\begin{equation}\label{eq:posterior_approx}
    P(\p \, | \, \A, \X) \approx \N(\p; \hat{\m}^{\p}, \hat{\s}^{\p}) \, , \, \forall \, \p \in \T \, .
\end{equation}
LA involves a second-order Taylor expansion around the Maximum A Posteriori estimate~(MAP) of the right-hand side of \cref{eq:posterior}. We compute this estimate using Automatic Differentiation~(AD), a gradient-based method that, in our implementation, employs the Adam optimizer to iteratively evaluate derivatives of the log-posterior. One might question the validity of this approximation in a potentially multimodal landscape. However, it is important to recognize that the existence of local minima is a problem independent of the approximation method employed. Nonetheless, in our model class -- multilevel networks of log-convex likelihood functions connected by log-convex link functions -- this issue is less likely to occur.

The MAP estimate found with AD also constitutes the mean $\hat{\m}^{\p}$ of $P(\p \, | \, \A, \X)$. To go beyond point estimates and quantify uncertainty, one can further estimate the covariance matrix $\hat{\s}^{\p}$, which is given by the inverted Hessian around the MAP:
\begin{equation}\label{eq:sigma_hat}
    \hat{\s}^{\p} \approx \Big[- \frac{\partial^2 P(\p \, | \, \A, \X)}{\partial \p}(\hat{\m}^{\p}) \Big]^{-1} \, .
\end{equation}

Other inference methods can be employed to approximate Gaussian distributions, such as VI. However, in such situations, utilizing AD directly might not be feasible due to the involvement of uncertain expectations in the optimization cost function. On the other hand, LA naturally combines with AD, providing a flexible and efficient inference procedure.

\subsection*{Parameter interpretation}\setcurrentname{Parameter interpretation}\label{subsec:parameter_interpretation}
We approximate the posterior distributions of the latent variables using Gaussian distributions, as outlined in \cref{eq:posterior_approx}. Consequently, all our estimated parameters belong to the real space. Although this approach is advantageous for developing an efficient and automated inference method, practitioners may desire different variable domains to enhance interpretability. In some instances, achieving this transformation is straightforward, involving the application of the probability transformation rule to obtain a distribution for the transformed variable within the desired constrained support. For example, if we are interested in expressing $\bar{\U} := \exp(\hat{\U}) \in \mathbb{R}_{>0}^{N \times K}$, we can simply employ the $\Lognormal(\bar{\U}; \hat{\m}^{\U}, \hat{\s}^{\U})$ distribution. Similarly, when seeking $\bar{\U} := \logistic(\hat{\U}) \in (0,1)^{N \times K}$, we can just compute the $\Logitnormal(\bar{\U}; \hat{\m}^{\U}, \hat{\s}^{\U})$. 

However, certain functions lack closed-form transformations. For instance, obtaining a probabilistic interpretation of the mixed-memberships of nodes requires applying the $\softmax$ function to each row of the matrices $\U$ and $\V$, which is not a bijective function. To address this challenge, our framework employs the Laplace Matching~(LM)~\cite{hobbhahn2021laplace} to approximate the distributions of such transformations. This technique yields a bidirectional, closed-form mapping between the parameters of the Gaussian distribution and those of the approximated transformed distribution. In this scenario, we can derive:
\begin{align}\label{eq:laplace_matching_example}
    \bar{\U}_{\id}:= &\softmax(\hat{\U}_{\id}) \, , \bar{U}_{ik} \in [0,1] \, \text{and} \, \sum_{k=1}^K \bar{U}_{ik} = 1 \nonumber \\
    &\quad \, \, \text{with} \, P(\bar{\U}_{\id}) = \Dir(\bar{\U}_{\id}; \hat{\al}^{\U}_{\id}) \, ,
\end{align}
where $\hat{\al}^{\U}_{\id}$ is a $K$-dimensional vector obtained with LM, whose entries are described as:
\begin{equation}
 \hat{\alpha}^{U}_{ik} = \frac{1}{\hat{\Sigma}^{U}_{ikk}}\Big(1 - \frac{2}{K} + \frac{\exp(\hat{\mu}^{U}_{ik})}{K^2} \sum_{l = 1}^{K} \exp(\hat{\mu}^{U}_{il}) \Big) \, .
\end{equation}

This approach is theoretically grounded and enables us to provide closed-form posterior distributions for the latent variables across a diverse range of domains. Consequently, it consistently allows for the estimation of uncertainties and other relevant statistical measures. Nonetheless, \model\ can also be utilized for the sole purpose of determining point estimates of the latent variables, which are essentially given by the MAP estimates. In such scenarios, it remains feasible to map these point estimates to different supports by applying any desired function, without worrying about the transformation process. Although this approach lacks full posterior distributions, it significantly simplifies the inference process by avoiding the computation of the Hessian. The choice between these two approaches should be guided by the specific application under study.

\section*{Results}
We demonstrate our method on both synthetic and real-world datasets, presenting a comprehensive analysis through quantitative and qualitative findings. Further explanations about the data generation and pre-processing procedures can be found in the Supporting Information, which also includes additional results. The settings used to run our experiments and the choice of the hyperparameters are also described in the Supporting Information (see \cref{sec:model_setting}). The code implementation of \model\ is accessible at: \href{https://github.com/mcontisc/PIHAM}{https://github.com/mcontisc/PIHAM}.

\subsection*{Simulation study}
\paragraph*{\normalfont \textbf{Comparison with existing methods in a homogeneous scenario.}} \setcurrentname{Comparison with existing methods in a homogeneous scenario}\label{subsec:synthetic_comparison}
We first investigate the behavior of our model in a simpler and common scenario, characterized by attributed multilayer networks with nonnegative discrete weights and one categorical node attribute. This represents the most general case addressed by existing methods, which are specifically designed for homogeneous settings, where there is only one attribute and one data type. For comparison, we use \mtcov~\cite{contisciani2020community}, a probabilistic model that assumes overlapping communities as the main mechanism governing both interactions and node attributes. In contrast to \model, \mtcov\ is tailor-made to handle categorical attributes and nonnegative discrete weights. Additionally, it employs an EM algorithm, with closed-form derivations for parameters inference strongly relying on the data type, making \mtcov\ a bespoke solution compared to the more general framework proposed by \model. The results of this comparison are depicted in \cref{fig:synthetic_1} of the Supporting Information, accompanied by additional details about the data generation and experiment settings. In principle, we expect \mtcov\ to exhibit better performance in this specific scenario due to its tailored development for such data and also its generative process aligning closely with the mechanism underlying the synthetic data. Nonetheless, despite the generality of our approach, we observe that \model\ achieves comparable performance to \mtcov\ in link and attribute prediction, as well as community detection, especially in scenarios involving denser networks. These results collectively show that \model\ is a valid approach even in less heterogeneous scenarios, as it can compete effectively with bespoke existing methods.

\paragraph*{\normalfont \textbf{Validation on heterogeneous data.}}\setcurrentname{Validation on heterogeneous data}\label{subsec:synthetic_validation}
Having demonstrated that \model\ performs comparably well to existing methods for attributed multilayer networks, we now demonstrate its behavior on more complex data containing heterogeneous information. To the best of our knowledge, this is the first probabilistic generative model designed to handle and perform inference on such data, and as a result, a comparative analysis is currently unavailable. Additionally, due to the absence of alternative benchmarks for data generation, we validate the performance of our method on synthetic data generated using the model introduced in this work.

\begin{figure*}[ht!]
\centering
\includegraphics[width=0.9\textwidth]{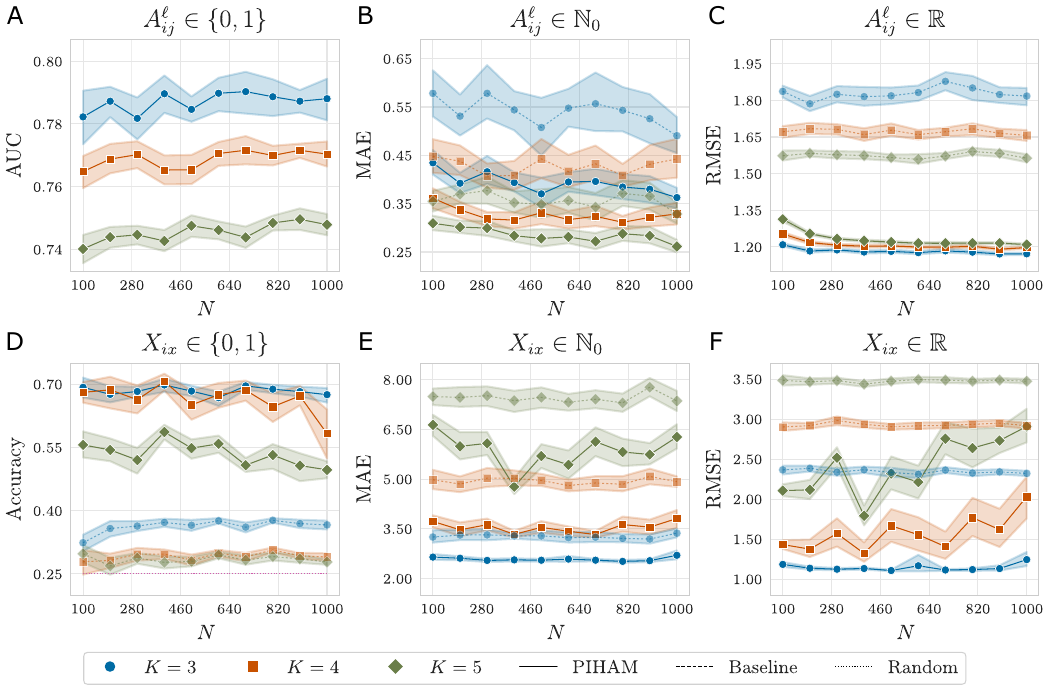}
\caption{Prediction performance on synthetic data. We analyze synthetic attributed multilayer networks with $L=3$ heterogeneous layers (one with binary interactions (A), one with nonnegative discrete weights (B), and one with real values(C)), three node covariates (one categorical with $Z=4$ categories (D), one representing nonnegative discrete values (E), and one involving real values (F)), varying number of nodes $N$, and diverse number of overlapping communities $K$. We employ a $5$-fold cross-validation procedure and plot averages and confidence intervals over $20$ independent samples. The prediction performances are measured with different metrics according to the data type: Area Under the receiver-operator Curve (AUC) for binary interactions (A), the Maximum Absolute Error (MAE) for nonnegative discrete values (B, E), the Root Mean Squared Error (RMSE) for real values (C, F), and accuracy for categorical attributes (D). The baselines are given by the predictions obtained from either the average or the maximum frequency in the training set. For the categorical attribute, we also include the uniform random probability over $Z$, and for the AUC, the baseline corresponds to the random choice~$0.5$. Overall, \model\ outperforms the baselines significantly for each type of information.}
\label{fig:prediction_synthetic}
\end{figure*}

We analyze attributed multilayer networks with $L=3$ heterogeneous layers: one with binary interactions, one with nonnegative discrete weights, and one with real values. In addition, each node is associated with three covariates: one categorical with $Z=4$ categories, one representing nonnegative discrete values, and one involving real values. To generate these networks, we initially draw the latent variables $\T=(\U, \V, \w, \h)$ from Gaussian distributions with specified hyperparameters. Subsequently, we generate $\A$ and $\X$ according to the data types, following \cref{eq:likelihood_network} and \cref{eq:likelihood_metadata}. Our analysis spans networks with varying number of nodes $N \in \{100, 200, \dots, 1000\}$ and diverse number of overlapping communities $K \in \{3, 4, 5\}$. Additional details on the generation process can be found in \cref{sec:exp_syn2} of the Supporting Information. 

We assess the effectiveness of \model\ by testing its prediction performance. To this end, we adopt a $5$-fold cross-validation procedure, where we estimate the model's parameters on the training set and subsequently evaluate its prediction performance on the test set (see the Supporting Information for details). The presence of heterogeneous information complicates the measurement of goodness of fit, as distinct data types impose different constraints and domains. To address this complexity, we employ different metrics tailored to assess the prediction performance of each type of information. Specifically, we use the Area Under the receiver-operator Curve (AUC) for binary interactions, the Maximum Absolute Error (MAE) for nonnegative discrete values, the Root Mean Squared Error (RMSE) for real values, and the accuracy for categorical attributes. Further exploration to determine a unified metric could be a subject of future research.

\begin{figure*}[ht!]
\centering
\includegraphics[width=0.9\textwidth]{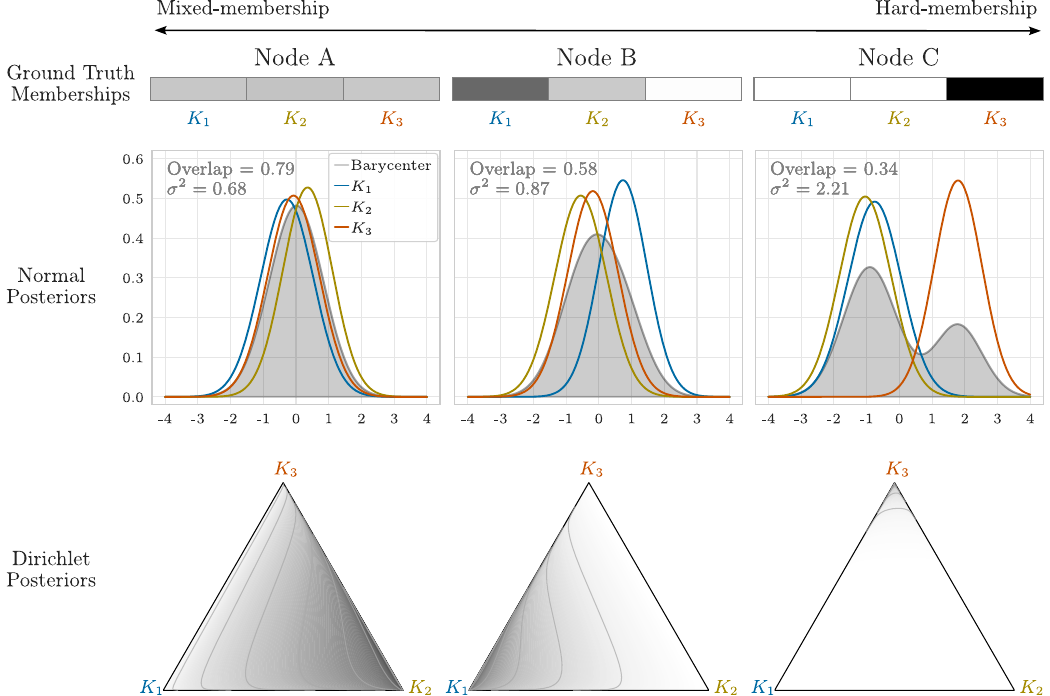}
\caption{Interpretation of posterior distributions in comparison with ground truth memberships. We analyze a synthetic attributed multilayer network with ground truth mixed-memberships represented as normalized vectors summing to $1$. In this case, $K=3$. (Top row) Ground truth membership vectors for three representative nodes: Node A displays extreme mixed-membership, Node B shows a slightly lower mixed-membership, and Node C exhibits hard-membership. (Middle row) Inferred posterior distributions $\hat{U}_{ik} \sim \N(\hat{U}_{ik};  \hat{\mu}^{U}_{ik}, (\hat{\sigma}^{U}_{ik})^{2})$, where different colors represent distinct communities, and the distribution in gray consists of the $L_2$-barycenter distribution. Overlap is the average of the area of overlap between every pair of distributions, and $\sigma^2$ is the variance of the barycenter distribution. (Bottom row) Transformed posterior distributions into the simplex space using the LM technique and employing Dirichlet distributions. The inferred node memberships reflect the ground truth behavior, as evidenced by the trends of Overlap and $\sigma^2$, which align with the decreasing degree of true mixed-membership. Additionally, the Dirichlet transformation provides a more straightforward interpretation, further supporting this conclusion.}
\label{fig:interpretability_synthetic}
\end{figure*}

The results are illustrated in \cref{fig:prediction_synthetic}, where the performance of \model\ is compared against baselines given by the predictions obtained from either the average or the maximum frequency in the training set. For the categorical attribute, we also include the uniform random probability over $Z$, and for the AUC, the baseline corresponds to the random choice~$0.5$. Overall, \model\ outperforms the baselines significantly for each type of information, with performance slightly decreasing as $K$ increases. This is somewhat expected, considering the increased complexity of the scenarios. On the other hand, the performance remains consistent across varying values of $N$, indicating the robustness of our method and its suitability for larger networks. 

\paragraph*{\normalfont \textbf{Interpretation of posterior estimates.}}
We have showcased the prediction performance of \model\ across diverse synthetic datasets, and we now delve into the qualitative insights that can be extracted from the inferred parameters. In particular, we focus on the membership matrix $\U$. For this purpose, we examine the results obtained through the analysis of the synthetic data used in the section \nameref{subsec:synthetic_comparison}, where ground truth mixed-memberships are represented as normalized vectors summing to $1$. This scenario is particularly relevant for illustrating an example where the desired parameter space, defined by the simplex, differs from the inferred one existing in real-space.

To ease visualizations, we investigate a randomly selected network and focus on three representative nodes with distinct ground truth memberships: Node A has extreme mixed-membership, Node B slightly less mixed-membership, and Node C exhibits hard-membership. The results are depicted in \cref{fig:interpretability_synthetic}, with the top row displaying the ground truth membership vectors for these representative nodes. In the middle row, we plot the inferred posterior distributions $\hat{U}_{ik} \sim \N(\hat{U}_{ik}; \hat{\mu}^{U}_{ik}, (\hat{\sigma}^{U}_{ik})^{2})$, where different colors represent distinct communities (in this case, $K=3$). Through a comparative analysis of the three distributions for each node, we can gain insights into the nodes' behaviors: Node A exhibits greater overlap among the three distributions, Node B shows a slighter shift toward $K_1$,  while Node C distinctly aligns more with community $K_3$. This preliminary investigation leads to the conclusion that the inferred communities reflect the ground truth behaviors. However, interpreting such patterns can be challenging, if not unfeasible, especially when dealing with large datasets. To address this issue, we quantitatively compute the area of overlap between every pair of distributions for each node and then calculate the average. For this purpose, we use the implementation proposed in~\cite{wand2011mean} and we name this measure as \textit{Overlap}. This metric ranges from $0$ (indicating no overlap) to $1$ (representing perfect matching between the distributions). Notably, the overlap decreases as we move from Node A to Node C, in line with the decreasing degree of mixed-membership.

Computing the Overlap for many communities can be computationally expensive due to the need to calculate all pairwise combinations. As an alternative solution, we suggest utilizing the $L_2$-barycenter distribution, which essentially represents a weighted average of the node-community distributions~\cite{benamou2015iterative, coz2023barycenter}. We show the barycenter distributions in gray in the second row of \cref{fig:interpretability_synthetic}. This approach allows focusing on a single distribution per node, instead of $K$ different ones. To quantify this distribution, we calculate its variance ($\sigma^2$), where higher values indicate nodes with harder memberships, as the barycenter is more spread due to the individual distributions being more distant from each other. Conversely, lower variance suggests more overlap among the distributions, indicating a more mixed-membership scenario. We observe that $\sigma^2$ increases as we decrease the degree of mixed-membership, a trend consistent with that of the Overlap. Further details on the barycenter distribution and the metrics are provided in \cref{sec:syn_interpretation} of the Supporting Information.

To facilitate interpretability, a practitioner may desire to work within the simplex space. This also reflects the ground truth parameter space, as opposed to the normal posterior distributions. As discussed in the section \nameref{subsec:parameter_interpretation}, \model\ employs the LM technique. This has the capability to transform in a principled way every membership vector $\hat{\U}_{\id}$ into the simplex space using Dirichlet distributions. The outcomes of this transformation are depicted in the bottom row of \cref{fig:interpretability_synthetic}. By investigating these plots, it becomes even more apparent how the inferred memberships closely resemble the ground truth: the Dirichlet distributions gradually concentrate more towards a specific corner ($K_1$ for Node B and $K_3$ for Node C), instead of spreading across the entire area (as observed for Node A).

With this example, we presented a range of solutions for interpreting the posterior distributions associated with the inferred node memberships. These options are not exhaustive, and other approaches may also be considered. For instance, a practitioner might focus solely on analyzing the point estimates for the sake of facilitating comparisons with the ground truth. In such cases, as discussed in the section \nameref{subsec:parameter_interpretation}, two procedures can be employed: i) applying a transformation to the point estimates, such as $\softmax$, to align them with the ground truth space, or ii) using a sufficient statistic of the posterior distribution, where the mean of the Dirichlet distribution is a suitable option. The choice between these various approaches should be guided by the specific application under study, and the provided example serves as just one illustration.

\subsection*{Analysis of a social support network of a rural Indian village}
We now turn our attention to the analysis of a real-world dataset describing a social support network within a village in Tamil Nadu, India, referred to as ``A\underline{l}ak\={a}puram''~\cite{power2015building, power2017social}. The data were collected in 2013 through surveys, in which adult residents were asked to nominate individuals who provided various types of support, such as running errands, offering advice, and lending cash or household items. Additionally, several attributes were gathered, encompassing information like gender, age, and caste, among others. The pre-processing of the dataset is described in \cref{sec:real_si} of the Supporting Information. The resulting heterogeneous attributed multilayer network comprises $N = 419$ nodes, $L=7$ layers, and $P=3$ node attributes. The initial six layers depict directed binary social support interactions among individuals, with average degree ranging from $1.8$ to $4.2$. The seventh, instead, contains information that is proportional to the geographical distance between individuals’ households. The adjacency tensor is then represented as $\A = \{\Abin \in \{0, 1\}^{N \times N} \, \forall \ell \in [1, 6],  \Adist \in \mathbb{R_+}^{N \times N} \}$. As node covariates, we consider the caste attribute with $Z_{caste} = 14$ categories, the religion attribute with $Z_{religion} = 3$ categories, and the attribute representing the years of education, that is $\xedu \in \mathbb{N}_0^{N}$. Ethnographic work and earlier analyses~\cite{power2017social, power2018building} suggest that these attributes play an important role in how villagers relate to one another, with certain relationships being more strongly structured by these identities than others.

\paragraph*{\normalfont \textbf{Inference, prediction performance, and goodness of fit.}}
We describe the likelihood of the real-world heterogeneous attributed multilayer network according to \cref{eq:likelihood_network} and \cref{eq:likelihood_metadata}, customized to suit the data types under examination. In particular, we employ Bernoulli distributions for the binary layers $[\Abin]_{\ell \in [1, 6]}$ and Gaussian distributions for the distance layer $\Adist$. Moreover, we characterize the attributes caste $\xcaste$ and religion $\xrel$ using Categorical distributions, and model the covariate $\xedu$ with a Poisson distribution. The choice of the model hyperparameters and the algorithmic settings used in our experiments are described in \cref{sec:model_setting} of the Supplementary Information. 

Similarly to many real-world datasets, we lack the information about the true parameters underlying the network, including the node memberships. Hence, to determine the number of communities $K$, we employ a $5$-fold cross-validation procedure for $K \in [1, 10]$ and select the value that exhibits the optimal performance. Detailed results are displayed in \cref{tab:results_cv} of the Supporting Information. We set $K=6$ as it achieves the best performance across the majority of prediction metrics. In fact, selecting a single metric to summarize and evaluate results in a heterogeneous setting is nontrivial, as discussed in the section \nameref{subsec:synthetic_validation}. The results in \cref{tab:results_cv} additionally validate \model's performance in inference tasks like edge and covariate prediction. Overall, our method demonstrates robust outcomes with the chosen fixed value of $K$ and consistently outperforms the baselines, which are omitted for brevity.

We further evaluate our model's goodness of fit through a posterior-predictive assessment~\cite{gelman1996posterior, gelman2013bayesian}, comparing the input data to synthetic data generated by the fitted model. A well-fitted model should produce synthetic data that closely resemble the original input. To accurately assess performance, we test whether two samples from the posterior-predictive distribution are generally more, equally, or less distant from each other than a sample from the posterior-predictive distribution compared to the input data~\cite{gelman1996posterior}. We measure the distance using different metrics depending on the data type, and the results are shown in \cref{fig:posterior_predictive}. The discrepancies between synthetic data samples consistently exceed those between the observed data and synthetic samples, indicating that \model\, provides a good fit for the data.

\begin{figure*}[ht!]
\centering
\includegraphics[width=\textwidth]{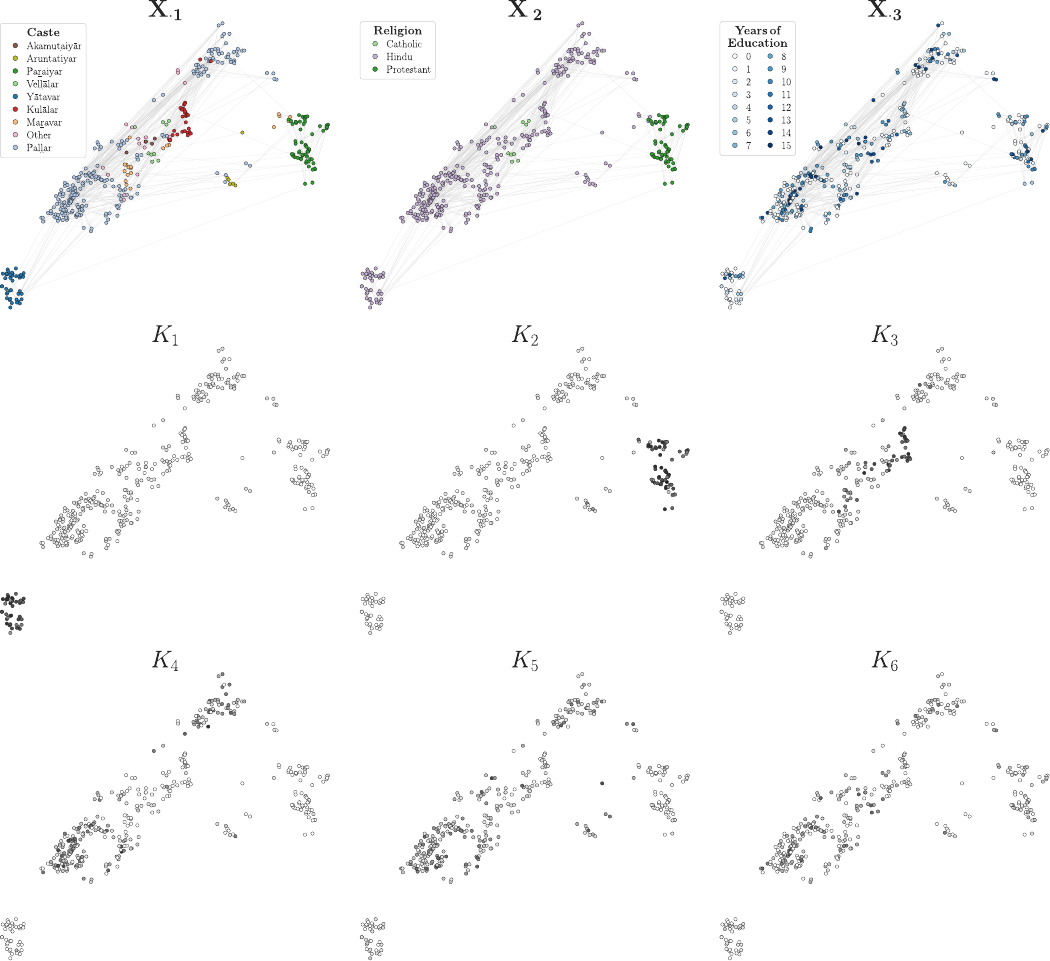}
\caption{Inference of overlapping communities in a social support network. We analyze a real-world heterogeneous attributed multilayer network, which was collected in 2013 through surveys in the Indian village. This network comprises six binary layers representing directed social support interactions among individuals, alongside an additional layer reflecting information proportional to the distance between individuals’ households. (Top row) As node covariates, we consider caste $\xcaste$, religion $\xrel$, and years of education $\xedu$. For privacy reasons, nodes belonging to castes with fewer than five individuals are aggregated into an ``Other'' category. Moreover, the displayed interactions refer only to the first layer (talk about important matters) to enhance clarity in visualization. (Middle-Bottom rows) We display the MAP estimates of the out-going communities inferred by \model. For easier interpretation, we apply a $\softmax$ transformation to the MAP estimates of the membership vectors, and darker values in the grayscale indicate higher values in the membership vector $\hat{\U}_{\id}$. The position of the nodes reflects the geographical distance between individuals' households. In summary, the inferred communities do not exclusively align with a single type of information. Rather, \model\ incorporates all input information to infer partitions that effectively integrate them in a meaningful way.}
\label{fig:communities_real}
\end{figure*}

\paragraph*{\normalfont \textbf{Qualitative interpretation of the inferred parameters.}}
We now shift our attention to analyze the results qualitatively, specifically focusing on the inferred communities. For easier interpretation, we apply a $\softmax$ transformation to the MAP estimates $\hat{\m}^{\U}_{\id}$, allowing us to treat node memberships as probabilities. Opting for the $\softmax$ over the mean of the posterior Dirichlet distributions is primarily for visualization purposes, as it results in slightly less mixed-memberships, thereby improving clarity. The middle and bottom rows of \cref{fig:communities_real} depict the inferred out-going communities $\hat{\U}_{\id}$, where darker values in the grayscale indicate higher values in the membership vector. In addition, the top row of \cref{fig:communities_real} displays the node attributes included in our analysis. Note also that the nodes' position reflects the geographical distance between individuals' households, and the depicted interactions refer to the first layer (talk about important matters). A full representation of the six binary layers is shown in \cref{fig:ala_layers} of the Supporting Information.

Upon initial examination, we observe a correspondence between various detected communities and the covariate information. For instance, the first and second communities predominantly consist of nodes belonging to the Y$\mathrm{\bar{a}}$tavar and Pa$\mathrm{\underbar{r}}$aiyar castes, respectively. Similarly, $K_3$ comprises nodes from the Kul$\mathrm{\bar{a}}$lar and Ma$\mathrm{\underbar{r}}$avar castes. This observation is supported by the inferred $K \times Z_{caste}$-dimensional matrix $\hat{\boldsymbol{H}}_{\boldsymbol{\cdot1}}$ (see \cref{fig:ala_Hcaste} in the Supporting Information), which explains the contributions of each caste category to the formation of the $k$-th community. Furthermore, the affinity tensor $\hat{\w}$ (see \cref{fig:ala_W} in the Supporting Information) suggests that these communities have an assortative structure, where nodes tend to interact more with individuals belonging to the same community as with those from different communities. This pattern reflects a typical behavior in social networks~\cite{newman2002assortative}. Additionally, note that these communities contain nodes that are geographically close to each other and, in some cases, very distant from the majority.

In contrast to the first three, communities $K_4$, $K_5$, and $K_6$ are more nuanced. In fact, they are predominantly comprised of nodes from the Pal$\d{}$l$\d{}$ar caste, which, however, is also the most represented caste in the dataset. Despite that, we observe some differences by examining other parameters. For instance, $K_4$ exhibits a strong assortative community structure, contrasting with the less structured nature of $K_5$ and $K_6$. This suggests that interactions play a more relevant role than attributes in determining the memberships of $K_4$. On the other hand, the attribute $\xedu$ seems to play a bigger role in determining $K_6$, which includes nodes with more years of education. This correlation is depicted in \cref{fig:ala_Hedu} of the Supporting Information, where the posterior distribution $\mathcal{N}(\hat{H}_{63}; \hat{\mu}^{H}_{63}, (\hat{\sigma}^{H}_{63})^{2})$ of education years in $K_6$ significantly differs and is distant from the others.

By looking at the affinity matrices of the seven layers in \cref{fig:ala_W}, we see how layers have predominantly an assortative structure, but show also variations for certain layers. For instance, $L_{2}$ (help finding a job) has few non-zero diagonal values, suggesting that this type of support is one for which people must sometimes seek out others in different communities. In particular, $L_{7}$, corresponding to the geographical distance between nodes, has several off-diagonal entries, particularly for communities $K_4$, $K_5$, and $K_6$, suggesting a weakened effect for physical proximity for those communities.

Taken together, these findings suggest that the inferred communities do not solely correlate with one type of information, which may be the most dominant. Instead, \model\ utilizes all the input information to infer partitions that effectively integrate all of them in a meaningful manner. In addition, the inferred affinity matrices illustrate how different layers can exhibit different community structures, a diversity that can be captured by our model.

\section*{Discussion}
In this work, we have introduced \model, a probabilistic generative model designed to perform inference in heterogeneous and attributed multilayer networks. A significant feature of our approach is its flexibility to accommodate any combination of the input data, made possible through the use of Laplace approximations and automatic differentiation methods, which avoid the need for explicit derivations. However, it is important to note that having a method capable of handling complex network datasets does not automatically ensure the quality of the input data. For example, if only certain attributes are relevant for explaining the networked dataset, or if only specific layers contain valuable information for the task, adding unnecessary information could be detrimental. Practitioners must carefully assess which information is useful based on their specific objectives. Alternatively, model selection tests, such as the cross-validation routine demonstrated in the manuscript, can be used to determine the optimal combination of layers and attributes.

When compared to other methods tailored for scenarios with only one type of attribute and interaction, \model\ demonstrates comparable performance in prediction and community detection tasks, despite its broader formulation. Moreover, our approach significantly outperforms baseline metrics in more complex settings characterized by various attribute and interaction types, where existing methods for comparison are lacking. 
Additionally, \model\ employs a Bayesian framework, enabling the estimation of posterior distributions, rather than only providing point estimates for the parameters. And, through the use of the Laplace matching technique, it maps these posterior distributions to various desired domains in a theoretically sound manner, facilitating interpretation.

While \model\ constitutes a principled and flexible method to analyze heterogeneous and attributed multilayer networks, several questions remain unanswered. For example, determining the most appropriate metric for summarizing prediction performance in heterogeneous scenarios, where information spans different spaces, is not straightforward. This aspect also influences the selection of the optimal model during cross-validation procedures. While we have provided explanations for our choices, we acknowledge that this remains an open question. 
Similarly, when dealing with many communities, summarizing posterior distributions becomes challenging due to computational constraints. We addressed this issue by employing $L_2$-barycenter distributions and proposing their variance to guide interpretation. Nevertheless, we believe there is still considerable room for improvement and exploration in this area. 
Moreover, we have considered here mixed-memberships, but in certain data-scarce scenarios hard-membership approaches with fewer parameters could be better suited. Future work should consider how to flexibly drive parameters' inference towards mixed or hard memberships, based on the input data.
Our method could be further extended to accommodate distinct community-covariate contributions by integrating two separate $\h$ matrices for both in-coming and out-going communities, respectively. This modification will offer clearer insights into how covariates influence the partitions, especially when discrepancies arise between in-coming and out-going communities. Moreover, we treated the number of communities $K$ as given, and for real-world data it was selected via cross-validation. While this procedure is grounded, it can be computationally expensive. An alternative approach would be to treat $K$ as a model parameter and infer it directly from the data. Lastly, it would be interesting to expand this framework to incorporate higher-order interactions, an emerging area that has shown relevance in describing real-world data~\cite{badalyan2023hypergraphs}.

In summary, \model\ offers a flexible and effective approach for modeling heterogeneous and attributed multilayer networks, which arguably better captures the complexity of real-world data, enhancing our capacity to understand and analyze the organization of real-world systems.

\bibliographystyle{ScienceAdvances}
\bibliography{bibliography}

\vspace{5mm}
\noindent \textbf{Acknowledgements:} M.C. and C.D.B. were supported by the Cyber Valley Research Fund.
M.C. acknowledges support from the International Max Planck Research School for Intelligent Systems (IMPRS-IS).
M.C., C.D.B., and E.A.P. were supported by a UKRI Economic and Social Research Council Research Methods Development Grant (ES/V006495/1).
M.H. and P.H. acknowledge support from the European Research Council through ERC CoG Action 101123955 ANUBIS; the DFG Cluster of Excellence “Machine Learning - New Perspectives for Science”, EXC 2064/1, project number 390727645; the German Federal Ministry of Education and Research (BMBF) through the Tübingen AI Center (FKZ: 01IS18039A); as well as funds from the Ministry of Science, Research and Arts of the State of Baden-Württemberg. \\
 \textbf{Author contributions:} M.C., M.H., P.H., and C.D.B. designed research; M.C., M.H., and C.D.B. performed research; M.C., E.A.P., and C.D.B. analyzed data; and M.C., M.H., E.A.P., P.H., and C.D.B. wrote the paper. \\
\textbf{Data availability}: Synthetic data used in the paper are explained in the Supplementary Material. Anonymized data of the social support network are available from the corresponding author E. A. Power, upon reasonable request.\\
\textbf{Code availability}: An open-source algorithmic implementation of the model is publicly available and can be found at \href{https://github.com/mcontisc/PIHAM}{https://github.com/mcontisc/PIHAM}.

\newcommand{\beginsupplement}{%
        \renewcommand{\thesection}{S\arabic{section}}
        \renewcommand{\theequation}{S\arabic{equation}}
	\renewcommand{\thetable}{S\arabic{table}}
	\renewcommand{\thefigure}{S\arabic{figure}}
 }

\clearpage
\beginsupplement
\onecolumngrid
\section*{{Supporting Information}}

\section{Modelling categorical node metadata}\label{sec:cat_attribute}
\cref{eq:pi} in the main text describes the general formulation of the expected value of an attribute $x$ for node $i$. However, when the attribute $x$ is categorical, the expression becomes more intricate because it has to account for the total number of attribute categories $Z$. In this case, $\boldsymbol{\pi_{ix}}(\T) = [\pi_{ixz}(\T)]_{z \in [1,Z]}$ and $\boldsymbol{H_{kx}}=[H_{kxz}]_{z \in [1,Z]}$ are $Z$-dimensional vectors. Within this framework, $H_{kxz}$ explains how much information from the $z$-th category of attribute $x$ is used to create the $k$-th community, and $\pi_{ixz}(\T)$ follows the modelling approach outlined in \cite{contisciani2020community}: 
\begin{equation} \label{eq:pi_cat}
\pi_{ixz}(\T) \approx \frac{1}{2} \sum_{k=1}^K (U_{ik} + V_{ik}) \, H_{kxz}  \, .
\end{equation}

\section{Model settings and hyperparameters choice}\label{sec:model_setting}
We employ \model\ consistently, maintaining the same configurations and hyperparameters across all experiments. The only variation lies in the choice of the likelihood function, customized to match the data types under examination. Specifically, we adopt Bernoulli distributions for binary information, Poisson distributions for nonnegative discrete data, Gaussian distributions for real values, and Categorical distributions for categorical data.

We set the prior distributions as standard normal distributions $\mathcal{N}(0,1)$, serving as shrinkage regularization. Indeed, due to the complexity of the data, the objective function may become non-identifiable, thus requiring the enforcement of concavity and differentiability. The selection of these priors accommodates this necessity. On the contrary, for parameter initialization, we opt for wider normal distributions $\mathcal{N}(0,9)$ to facilitate exploration of various initial points.

\model\ performs inference using the gradient-based method Automatic Differentiation, and in our implementation we employs the Adam optimizer to iteratively evaluate derivatives of the log-posterior. We set the learning rate of the optimizer equal to $0.5$ and run the optimization procedure for $2000$ iterations, maintaining a tolerance threshold of $10^{-8}$. Furthermore, we execute the algorithm $50$ times, each time starting from a different random initialization, and output the parameters corresponding to the realization with the highest objective function.

\section{Comparison with existing methods in a homogeneous scenario}
\subsection*{Data generation}
We construct directed attributed multilayer networks following a procedure similar to that described in~\cite{contisciani2020community}. Initially, we generate interactions using a multilayer mixed-membership stochastic block model~\cite{debacco2017community}. Subsequently, we assign node metadata, ensuring a $50\%$ match with the node communities, while the remaining ones are made randomly. We set a configuration with $N=500$ nodes, $L=2$ layers of which one being assortative and the other disassortative, a categorical attribute with $Z=3$ categories, and $K=3$ overlapping communities. Networks are generated with increasing average degrees $\avg \in \{10, 15, 20, \dots, 50\}$, producing $20$ independent samples for each $\avg$ value. To generate the membership matrices $\U$ and $\V$, we initially assign equal-size unmixed group memberships and then introduce the overlapping for $20\%$ of the nodes. The correlation between  $\U$ and $\V$ is set equal to $0.1$, with entries drawn from a Dirichlet distribution with parameter $\alpha=0.1$. The affinity matrix $\Wone$ exhibits an assortative block structure with main probabilities $p_1 = \avg K / N$ and secondary probabilities $p_2 = 0.1 \, p_1$. Conversely, the affinity matrix $\Wtwo$ is generated using a disassortative block structure with off-diagonal probabilities $p_1 = \avg K / N$ and diagonal probabilities $p_2 = 0.1 \, p_1$. Self-loops are removed, and sparsity is preserved. 

The resulting networks depict a simpler scenario characterized by homogeneous layers with nonnegative discrete weights and a single categorical attribute. Such scenario is crucial for testing our model against existing methods. 

\subsection*{Experiment details} 
For comparison, we use \mtcov~\cite{contisciani2020community}, a probabilistic model that assumes overlapping communities as the main mechanism governing both interactions and node attributes. This model is specifically tailored to handle categorical attributes and nonnegative discrete weights, and employs an EM algorithm for parameter inference. We run \mtcov\ $50$ times with different random initializations, maintaining the same tolerance as in our implementation. In addition, we set the maximum number of EM steps before termination at $500$, and the threshold for declaring convergence based on the consecutive updates respecting the tolerance equal to $15$. Lastly, we fix the scaling hyperparameter $\gamma=0.5$, reflecting the matching constraint imposed in the synthetic data generation. 

We assess the performance of \model\ and \mtcov\ in both prediction and community detection tasks. Specifically, we evaluate their predictive capabilities using a $5$-fold cross-validation routine, in which the dataset is split into five equal-size groups (folds), selected uniformly at random. The models are then trained on four of these folds (training set), covering $80\%$ of the triples $(i, j, \ell)$ and $80\%$ of the categorical vector entries, to learn their parameters. Next, we evaluate the models' performance on the held-out group (test set). This process is repeated five times by varying the test set, resulting in five trials per iteration. As performance metrics, we use the Area Under the Curve (AUC) for the edge prediction, which represents the probability that a randomly selected edge has a higher expected value than a randomly selected non-existing edge, and accuracy for covariate prediction. 

To evaluate the methods' performance in recovering communities, we first need to transform the inferred memberships to match the parameter space of the planted communities. Indeed, the ground truth mixed-memberships are represented as normalized vectors summing to $1$, whereas our inferred parameters belong to the real-space. As discussed in the section \nameref{subsec:parameter_space} of the main text, two approaches can be adopted for such alignment: i)~applying a $\softmax$ transformation to the point estimates $\hat{\m}^{\p}$, or ii)~employing the LM technique to obtain Dirichlet posterior distributions and utilizing a suitable statistic of these, where the mean serves as a viable option. In this experiment, we use both methods. For assessing performance, we utilize the Cosine Similarity (CS), a metric adept at capturing both hard and mixed-membership communities, ranging from $0$ (indicating no similarity) to $1$ (denoting perfect recovery). We compute the average cosine similarities of both membership matrices $\U$ and $\V$, and then average them across the nodes.

\begin{figure}
\centering
\includegraphics[width=\textwidth]{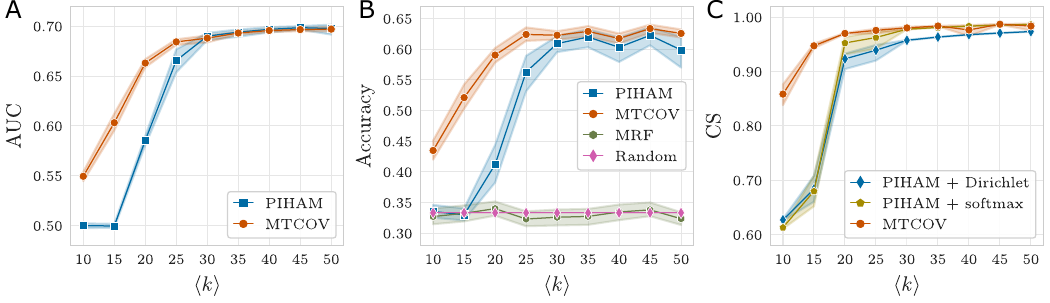}
\caption{\label{fig:synthetic_1}Prediction and community detection performance on synthetic data. We analyze synthetic attributed multilayer networks with $N=500$ nodes, $L=2$ layers (one being assortative and the other disassortative), a categorical attribute with $Z=3$ categories, $K=3$ overlapping communities, and increasing average degrees $\avg \in \{10, 15, 20, \dots, 50\}$. The results represent averages and confidence intervals over $20$ independent samples. For prediction tasks, we employ a $5$-fold cross-validation procedure. The evaluation metrics include (A) the AUC for edge prediction, with a baseline of $0.5$ corresponding to random choice, and (B) accuracy for covariate prediction. Here, MRF represents a baseline given by the predictions obtained from the maximum frequency in the training set, while Random denotes the uniform random probability over $Z$. (C) Community detection performance is assessed using Cosine Similarity (CS). As inferred point estimates, we consider both the mean of transformed Dirichlet posterior distributions and the $\softmax$ transformation of $\hat{\m}^{\p}$. Overall, \model\ exhibits comparable performance to \mtcov\ across all tasks despite its broader framework, especially in scenarios involving denser networks.}
\end{figure}

\subsection*{Results} 
The results of both models in prediction and community detection tasks are depicted in \cref{fig:synthetic_1}. While \mtcov\ is expected to exhibit better performance due to its close alignment with the generative process underlying the synthetic data, \model\ demonstrates comparable performance across all tasks despite its broader framework. This similarity is particularly notable in scenarios featuring denser networks. Indeed, our model, by treating all information equally, may face challenges in very sparse networks when relying only on a single covariate. Conversely, this is not an issue for \mtcov\ as it utilizes a linear combination of node and edge information, leveraging node covariates to address network sparsity.

An additional consideration is the choice of transformation used to compare the inferred communities with the planted ones. Both options yield similar results, as shown in \cref{fig:synthetic_1}C. The slight discrepancy between the two arises from the Dirichlet mean providing slightly more mixed-memberships.

Overall, these findings collectively suggest that \model\ remains a valid approach even in less heterogeneous scenarios, demonstrating its ability to effectively compete with ad-hoc existing methods.

\section{Validation on heterogeneous data}\label{sec:exp_syn2}
\subsection*{Data generation} 
We construct directed, heterogeneous, and attributed multilayer networks using the framework of \model. Initially, we draw the latent variables $\T=(\U, \V, \w, \h)$ from Gaussian distributions with specified hyperparameters, and then generate $\A$ and $\X$ according to the data types, following \crefrange{eq:likelihood_network}{eq:transformation_pi} in the main text. We configure the networks with $L=3$ heterogeneous layers: one with binary interactions, the second with nonnegative discrete weights, and the third with real values. Additionally, each node is associated with three covariates: one categorical with $Z=4$ categories, one with nonnegative discrete values, and the last with real values. Networks are generated with increasing number of nodes $N \in \{100, 200, \dots, 1000\}$, and varying number of overlapping communities $K \in \{3, 4, 5\}$. For each combination $(N, K)$, we generate $20$ different samples. To generate the membership matrices $\U$ and $\V$, we assign equal-size group memberships and draw the entries of the matrices from distributions with different means, according to the group the nodes belong to. Specifically, $u_{ik} \sim \mathcal{N}(2, 0.04)$ if $i$ is associated with group $k$, otherwise $u_{ik} \sim \mathcal{N}(-1, 0.04)$. Similarly, $v_{ik}\sim \mathcal{N}(2, 0.09)$ if $i$ is associated with group $k$, otherwise $v_{ik} \sim \mathcal{N}(-1, 0.09)$. The affinity tensor $\w$ exhibits an assortative block structure in each layer, with diagonal entries following normal distributions with zero mean and $\sigma=0.45$, and off-diagonal entries drawn from normal distributions with $\mu=-4$ and $\sigma=0.45$.  The community-covariate matrix $\h$ is set to maintain coherence between node interactions and covariates, avoiding additional noise in the data. For the categorical variable, $H_{kxz} \sim \mathcal{N}(0.5+k, 0.04)$ if $k=z$, otherwise $H_{kxz} \sim \mathcal{N}(0, 0.04)$. When $K=3$ or $K=5$, we set $H_{kx4} \sim \mathcal{N}(0.2, 0.04)$ or $H_{5xz} \sim \mathcal{N}(0.2, 0.04)$, respectively. The nonnegative discrete attribute is generated according to $H_{kx} \sim \mathcal{N}(1.5 \times \frac{k+2}{3} , 0.01)$, while the covariate with real values is constructed with $H_{kx} \sim \mathcal{N}(4 + (1 - k) \times 3 , 0.04)$.

The resulting networks represent a general scenario featuring heterogeneous interactions and node covariates, which is essential to validate our approach and demonstrate its flexibility.  

\subsection*{Experiment details} 
We assess the performance of \model\ by evaluating its predictive capabilities through a $5$-fold cross-validation routine. In this approach, the dataset is randomly divided into five equal-sized groups (folds), and the model is trained on four of them (the training set), which include $80\%$ of the triples $(i, j, \ell)$ and $80\%$ of the entries of each attribute vector, to learn its parameters. The performance of the model is then evaluated on the remaining fold (the test set). This process is repeated five times, each time with a different fold as the test set, resulting in five trials per iteration. For performance metrics, we use different measures depending on the type of information being evaluated. For binary interactions, we use AUC, which ranges from $0$ to $1$, with $0.5$ representing the random baseline. For nonnegative discrete data, we use the Maximum Absolute Error (MAE), and for real values, we use the Root Mean Squared Error (RMSE). In both cases, lower values indicate better performance. Additionally, for categorical attribute predictions, we use accuracy, which ranges from $0$ to $1$, with $1$ indicating perfect recovery.

\section{Interpretation of posterior estimates}\label{sec:syn_interpretation}
\subsection*{Experiment details} Interpreting posterior distributions can be challenging, especially with large datasets. To address this, we propose two different approaches to summarize the inferred results.

First, we employ a metric to quantify the area of overlap between distributions. We use the method proposed in~\cite{wand2011mean}, which defines the integrated absolute error (IAE) between two distributions as:
\begin{equation}
\text{IAE} = \int_{-\infty}^{\infty} \big| \, p_1(x) - p_2(x) \, \big| dx \, ,
\end{equation}
with $\text{IAE} \in [0, 2]$. When $p_1$ and $p_2$ completely overlap, the difference between them is a line in zero, and the IAE is $0$. Conversely, if there is no overlap, the IAE is $2$, which is the sum of the integrals of the two distributions. To normalize this metric to the range $[0,1]$, we define:
\begin{equation}
\text{Overlap} = 1-\frac{1}{2} \, \text{IAE} \, .
\end{equation}
In this case, an Overlap of $0$ indicates  no overlap between the distributions, while an Overlap of $1$ represents perfect matching. We compute this measure between every pair of distributions for each node and then calculate the average.

Computing the Overlap for many communities can be computationally expensive due to the need to evaluate all pairwise combinations. As an alternative, we use the $L_2$-barycenter distribution, which represents a weighted average of the node-community distributions~\cite{benamou2015iterative, coz2023barycenter}. This approach simplifies the problem by focusing on a single distribution per node instead of $K$ different ones. We calculate this distribution using the \texttt{POT Python package} \cite{flamary2021pot}, and we quantify it 
by computing its variance ($\sigma^2$) using the trapezoidal rule to approximate the integral. Higher values indicate nodes with harder memberships, as the barycenter is more spread due to the individual distributions being more distant from each other. Conversely, lower variance suggests more overlap among the distributions, indicating a more mixed-membership scenario. 

\section{Analysis of a social support network of a rural Indian village}\label{sec:real_si}
\subsection*{Data pre-processing}
\begin{figure}
\centering
\includegraphics[width=\textwidth]{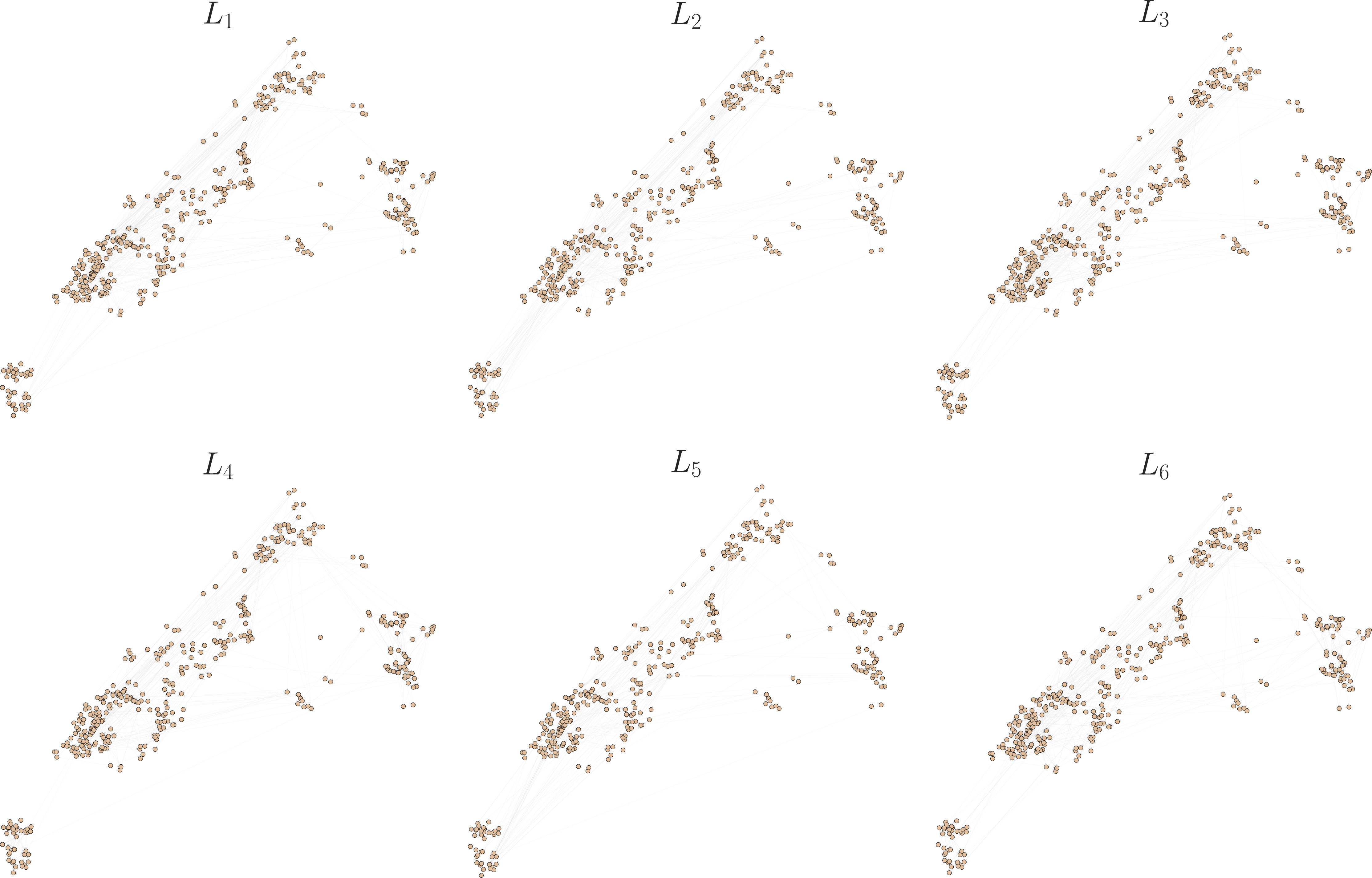}
\caption{Visual representation of the first six binary layers of the ``A\underline{l}ak\={a}puram'' social support network. Details on these layers, including the information encoded, the number of edges, and the average degree, are provided in \cref{tab:ala_stats}. The position of the nodes reflects the geographical distance between individuals’ households.}
\label{fig:ala_layers}
\end{figure}

We analyze a real-world dataset describing a social support network within a village in Tamil Nadu, India, referred to as ``A\underline{l}ak\={a}puram''~\cite{power2015building, power2017social}. The data were collected in 2013 through surveys, in which adult residents were asked to nominate individuals who provided various types of support. In our analysis, we consider six different binary support questions, each forming a layer in the network, and we exclude individuals without any of these interactions. Details on these layers, including the number of edges and the average degree, are provided in \cref{tab:ala_stats}, while a visual representation can be found in \cref{fig:ala_layers}. Additionally, we construct a seventh layer that incorporates the geographical distance between individuals' households. Specifically, we define the entries of this layer as $A_{ij}^{7} = \frac{1}{\sqrt{1+d_{ij}}}$, where $d_{ij}$ is the distance between the households of individuals $i$ and $j$, and we set $A_{ii}^{7}=0$. Note that higher values indicate closer proximity, while lower values represent greater distances. The resulting adjacency tensor is then represented as $\A = \{\Abin \in \{0, 1\}^{N \times N} \, \forall \ell \in [1, 6],  \Adist \in \mathbb{R_+}^{N \times N} \}$.

In addition, several attributes were collected, including information like gender, age, and caste, among others. For our analysis, we focus on caste, religion, and years of education, as ethnographic work and previous analyses~\cite{power2017social, power2018building} suggest these attributes significantly influence how villagers relate to one another. Specifically, caste is a categorical attribute with  $Z_{caste} = 14$ categories, religion has $Z_{religion} = 3$ categories, and years of education are represented as nonnegative discrete values $\xedu \in \mathbb{N}_0^{N}$. 

The resulting heterogeneous attributed multilayer network comprises $N = 419$ nodes, $L=7$ layers, and $P=3$ node attributes. 

\begin{table}[b]
\centering
\renewcommand{\arraystretch}{1.1}
\begin{tabular}{cccc}
Layer  & Description & E & $\langle k \rangle$ \\ 
\midrule
$L_1$  &   Talk about important matters  & $880$ & $4.2$ \\
$L_2$  &   Help finding a job  & $437$ & $2.1$ \\
$L_3$  &   Help with physical tasks  & $758$ & $3.6$ \\
$L_4$  &   Borrow household items from  & $876$ & $4.2$ \\
$L_5$  &   Ask for money  & $386$ & $1.8$ \\
$L_6$  &   Talk to for pleasure  & $824$ & $3.9$ \\
\bottomrule
\end{tabular}
\caption{\label{tab:ala_stats} Summary statistics for the first six binary layers of the ``A\underline{l}ak\={a}puram'' social support network. E denotes the number of edges, while $\langle k \rangle$ represents the average degree. A visual representation of these layers can be found in \cref{fig:ala_layers}.}
\end{table}

\subsection*{Results} 
To determine the number of communities $K$, we employ a $5$-fold cross-validation procedure for $K \in [1, 10]$ and select the value that exhibits the optimal performance. Similar to the synthetic experiments, for a given $K$, we train the model on four folds (training set), which include $80\%$ of the triples $(i, j, \ell)$ and $80\%$ of the entries of each attribute vector, to learn its parameters. The model's performance is then evaluated on the remaining fold (the test set). This process is repeated five times, each with a different fold as the test set, resulting in five trials per iteration. We evaluate prediction performance in the first six binary layers using the AUC, ranging from $0$ to $1$, with $0.5$ indicating the random baseline. For the seventh layer containing real values, we use the RMSE, where lower values indicate better performance. Additionally, we assess prediction performance for the attributes using accuracy for caste and religion, which ranges from $0$ to $1$ (with $1$ indicating perfect recovery), and the MAE for years of education, with lower values indicate better performance. The results are displayed in \cref{tab:results_cv}. In our experiments, we set $K=6$ as it achieves the best performance across most prediction metrics. Note that, summarizing and evaluating results in a heterogeneous setting using a single metric is challenging, as discussed in the section \nameref{subsec:synthetic_validation} of the main text. The results in \cref{tab:results_cv} also demonstrate that \model\ achieves robust outcomes with the chosen fixed value of $K$.

\begin{figure}
\centering
\includegraphics[width=\textwidth]{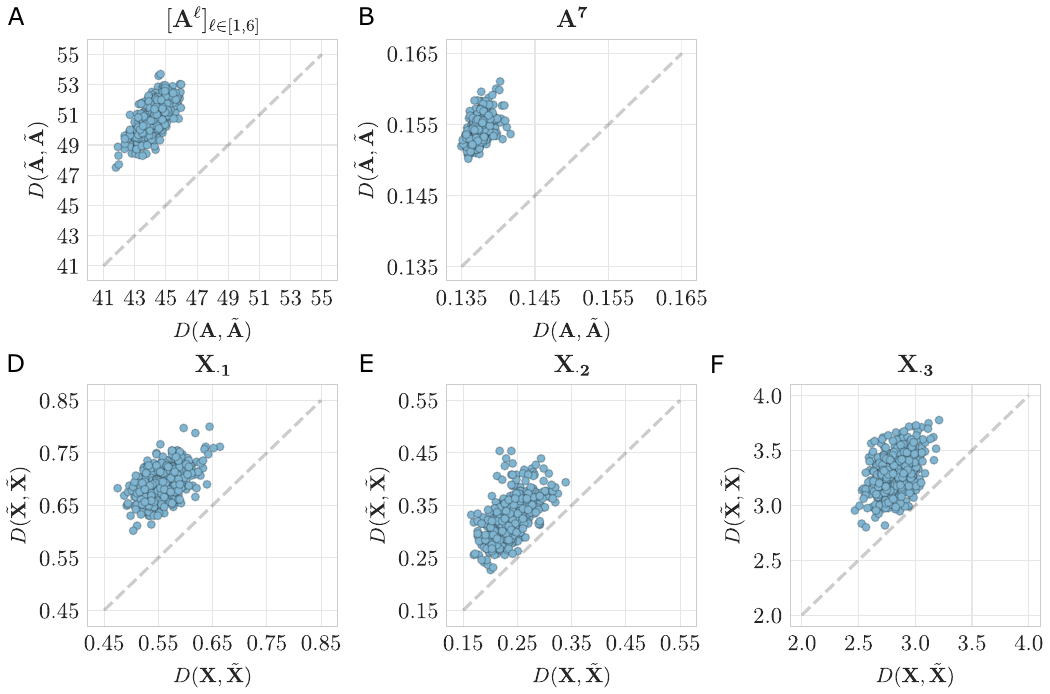}
\caption{Goodness of fit test for the ``A\underline{l}ak\={a}puram'' social support network. We generate $500$ independent synthetic datasets using the fitted model and compare them to both the input data and other synthetic samples. Each dot represents one posterior-predictive sample, showing the distance between this sample and the input data on the horizontal axis, and a randomly selected synthetic sample on the vertical axis. Performance is evaluated using different metrics $D$ depending on the data type: log-loss for binary interactions (A), RMSE for real values (B), $(1 - \text{accuracy})$ for categorical attributes (D, E), and MAE for nonnegative discrete values (F). The differences between synthetic data samples consistently exceed those between the observed data and synthetic samples, as all points lie above the diagonal, indicating that \model\ provides a good fit for the data.}
\label{fig:posterior_predictive}
\end{figure}

We further assess whether the model is a good fit to the data by conducting a posterior–predictive test~\cite{gelman1996posterior, gelman2013bayesian}, where we compare the input data $(\A, \X)$ to synthetic data $(\tilde{\A}, \tilde{\X})$ generated by the fitted model. A well-fitted model should produce synthetic data that closely match the original input. The posterior-predictive distribution is defined as
\begin{equation}
P(\tilde{\A}, \tilde{\X} \,|\, \A, \X) = \int_{\T} P(\tilde{\A}, \tilde{\X} \,|\, \T) \, P(\T \,|\, \A, \X) \, d\T  \, ,
\end{equation}
and samples from this distribution are obtained by first drawing $\T$ from the posterior distributions $\N(\p; \hat{\m}^{\p}, \hat{\s}^{\p})$, and then using these parameters to create new synthetic data. We apply the likelihood described in the previous subsection, fixing the standard deviation to $0.1$ for the Gaussian layer, and generate $500$ independent synthetic datasets. To evaluate performance, we test whether two samples from the posterior-predictive distribution are generally more, equally, or less distant from each other than a sample from the posterior-predictive distribution compared to the input data~\cite{gelman1996posterior}. We measure the distance using different metrics depending on the data type: log-loss for the six binary layers; RMSE for the seventh layer with real values, $(1 - \text{accuracy})$ for categorical attributes, and MAE for the attribute representing years of education. The results in \cref{fig:posterior_predictive} show that the discrepancies between synthetic data samples consistently exceed those between the observed data and synthetic samples, indicating that \model\, provides a good fit for the data.

\begin{table}[]
\centering
\renewcommand{\arraystretch}{1.1}
\begin{tabular}{c|c|c|c|c|c}
$K$  & AUC $([\Abin]_{\ell \in [1, 6]})$ & RMSE $(\Adist)$ & Accuracy $(\xcaste)$ & Accuracy $(\xrel)$ & MAE $(\xedu)$ \\ \hline
$1$  & $0.575 \pm 0.007$               & $0.096 \pm 0.003$      & $0.55 \pm 0.05$                & $0.84 \pm 0.04$                   & $4.3 \pm 0.3$                 \\
$2$  & $0.717 \pm 0.009$               & $0.096 \pm 0.003$      & $0.55 \pm 0.04$                & $0.84 \pm 0.04$                   & $4.3 \pm 0.2$                 \\
$3$  & $0.73 \pm 0.01$               & $0.096 \pm 0.003$      & $0.55 \pm 0.05$                & $0.84 \pm 0.04$                 & $\mathbf{4.3 \pm 0.1}$                 \\
$4$  & $\mathbf{0.77 \pm 0.01}$      & $0.093 \pm 0.003$      & $0.58 \pm 0.05$                & $0.84 \pm 0.03$                   & $4.4 \pm 0.1$                 \\
$5$  & $0.77 \pm 0.02$               & $0.088 \pm 0.005$      & $0.63 \pm 0.06$                & $0.87 \pm 0.06$                   & $4.4 \pm 0.1$                 \\
$6$  & $0.76 \pm 0.01$               & $\mathbf{0.086 \pm 0.003}$      & $\mathbf{0.70 \pm 0.07}$                & $\mathbf{0.87 \pm 0.02}$                   & $4.4 \pm 0.1$                 \\
$7$  & $0.74 \pm 0.01$               & $0.089 \pm 0.003$      & $0.63 \pm 0.04$                & $0.85 \pm 0.04$                   & $4.4 \pm 0.1$                 \\
$8$  & $0.73 \pm 0.01$               & $0.095 \pm 0.003$      & $0.27 \pm 0.06$                & $0.83 \pm 0.04$                   & $4.4 \pm 0.1$                 \\
$9$  & $0.72 \pm 0.01$               & $0.095 \pm 0.003$      & $0.20 \pm 0.04$                & $0.84 \pm 0.04$                   & $4.4 \pm 0.1$                 \\
$10$ & $0.72 \pm 0.01$               & $0.095 \pm 0.003$      & $0.2 \pm 0.2$                & $0.84 \pm 0.04$                   & $4.4 \pm 0.2$    \\           \hline
\end{tabular}
\caption{\label{tab:results_cv} Prediction performance on the ``A\underline{l}ak\={a}puram'' social support network. For a given number of communities $K$, we employ a $5$-fold cross-validation procedure and report averages and standard deviations over the five trials, each using a different fold as the test set. We evaluate prediction performance in the first six binary layers using the Area Under the Curve (AUC) and for the seventh layer containing real values using the Root Mean Squared Error (RMSE). Additionally, we assess prediction performance for the attributes using accuracy for caste~($\xcaste$) and religion ($\xrel$), and the Maximum Absolute Error (MAE) for years of education ($\xedu$), represented as nonnegative discrete values. The baselines are omitted for brevity. In our experiments, we set $K=6$ as it achieves the best performance across most prediction metrics, and overall, \model\ demonstrates robust outcomes with the chosen fixed value of $K$.}
\end{table}

\begin{figure}
\centering
\includegraphics[width=0.95\textwidth]{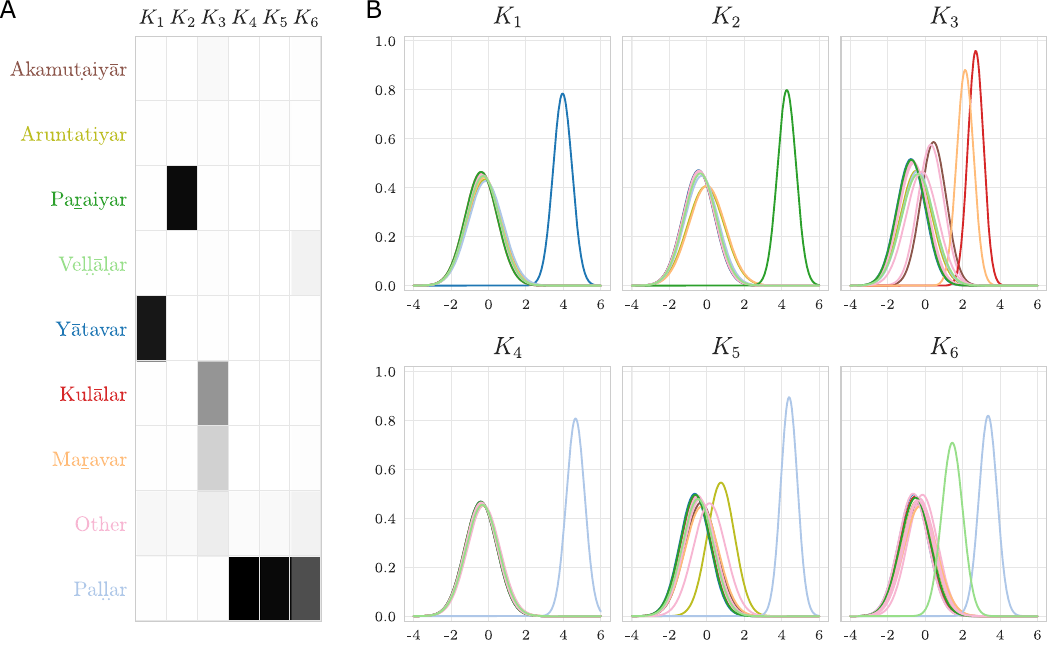}
\caption{Inference of the community-caste parameter in the ``A\underline{l}ak\={a}puram'' social support network. We display the inferred $K \times Z_{caste}$-dimensional matrix $\hat{\boldsymbol{H}}_{\boldsymbol{\cdot1}}$, which explains the contributions of each caste category to the formation of the $k$-th community. For privacy reasons, nodes belonging to castes with fewer than five individuals are aggregated into an “Other” category. (A) Transformation of the MAP estimates $\hat{\m}^{\h}_{\boldsymbol{k1}}$ inferred by \model\ using the $\softmax$ function. In this plot, the matrix is transposed, so that each column sums to $1$, and the y-axis lists the caste categories. (B) Inferred posterior distributions $\hat{H}_{k1z} \sim \mathcal{N}(\hat{H}_{k1z}; \hat{\mu}_{k1z}^{H}, (\hat{\sigma}_{k1z}^{H})^2)$, with different colors representing distinct caste categories. The first and second communities predominantly consist of nodes from to the Y$\mathrm{\bar{a}}$tavar and Pa$\mathrm{\underbar{r}}$aiyar castes, respectively. $K_3$ comprises nodes from the Kul$\mathrm{\bar{a}}$lar and Ma$\mathrm{\underbar{r}}$avar castes, while communities $K_4$, $K_5$, and $K_6$ are predominantly composed by nodes from the Pal$\d{}$l$\d{}$ar caste.}
\label{fig:ala_Hcaste}
\end{figure}

\begin{figure}
\centering
\includegraphics[width=0.45\textwidth]{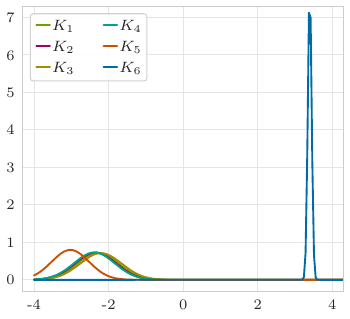}
\caption{Inference of the community-education parameter in the ``A\underline{l}ak\={a}puram'' social support network. We display the inferred posterior distributions $\hat{H}_{k3} \sim \mathcal{N}(\hat{H}_{k3}; \hat{\mu}_{k3}^{H}, (\hat{\sigma}_{k3}^{H})^2)$, which explain how the attribute related to years of education is distributed among the $K$ communities. The distribution of the attribute in $K_6$ has a notably higher mean compared to the distributions of the other communities, suggesting that $\xedu$ plays a significant role in determining this community.}
\label{fig:ala_Hedu}
\end{figure}

\begin{figure}
\centering
\includegraphics[width=\textwidth]{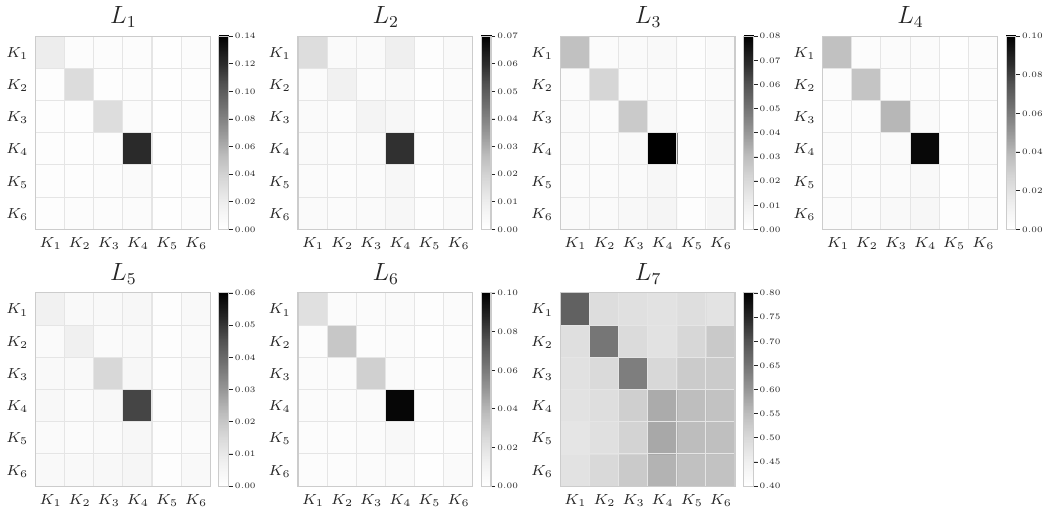}
\caption{Inference of the affinity tensor in the ``A\underline{l}ak\={a}puram'' social support network. We display the MAP estimates of the inferred $\hat{\w}$, which explain the edge density between different community pairs in the various layers. To improve visualization clarity, we apply a $\logistic$ transformation to the MAP estimates $[\hat{\mu}_{kq}^{\ell}]^{W}$. The different layers predominantly exhibit an assortative structure, with some variations for certain layers. For instance, $L_{2}$ (help finding a job) has few non-zero diagonal values, suggesting that this type of support sometimes requires seeking out individuals in different communities. Moreover, $L_{7}$, corresponding to the geographical distance between nodes, has several off-diagonal entries, particularly for communities $K_4$, $K_5$, and $K_6$, suggesting a weakened effect for physical proximity for those communities.}
\label{fig:ala_W}
\end{figure}

To provide a qualitatively interpretation of the inferred results, \cref{fig:communities_real} in the main text shows the inferred out-going communities $\hat{\U}$. Here, we present additional visualizations for other model parameters. In particular, \cref{fig:ala_Hcaste} displays the inferred $K \times Z_{caste}$-dimensional matrix $\hat{\boldsymbol{H}}_{\boldsymbol{\cdot1}}$, which explains the contributions of each caste category to the formation of the $k$-th community. Panel B presents the inferred posterior distributions $\hat{H}_{k1z} \sim \mathcal{N}(\hat{H}_{k1z}; \hat{\mu}_{k1z}^{H}, (\hat{\sigma}_{k1z}^{H})^2)$, with different colors representing distinct caste categories. Panel A, instead, shows the $\softmax$ transformation of the MAP estimates $\hat{\m}^{\h}_{\boldsymbol{k1}}$ for easier interpretation. Note that, the matrix is transposed in the plot, so that each column sums to $1$, and the y-axis lists the caste categories. From this figure, we observe that the first and second communities predominantly consist of nodes from to the Y$\mathrm{\bar{a}}$tavar and Pa$\mathrm{\underbar{r}}$aiyar castes, respectively. Similarly, $K_3$ comprises nodes from the Kul$\mathrm{\bar{a}}$lar and Ma$\mathrm{\underbar{r}}$avar castes, while communities $K_4$, $K_5$, and $K_6$ are predominantly composed by nodes from the Pal$\d{}$l$\d{}$ar caste, which is also the most represented caste in the dataset. Furthermore, \cref{fig:ala_Hedu} shows the inferred posterior distributions of the community-covariate vector related to years of education, $\hat{H}_{k3} \sim \mathcal{N}(\hat{H}_{k3}; \hat{\mu}_{k3}^{H}, (\hat{\sigma}_{k3}^{H})^2)$. From this figure, it is evident that this attribute plays a more significant role in determining $K_6$, as its distribution has a notably higher mean compared to the distributions of the other communities. Lastly, \cref{fig:ala_W} displays the affinity tensor $\hat{\w}$, which explains the edge density between different community pairs in the various layers. To improve visualization clarity, we apply a $\logistic$ transformation to the MAP estimates $[\hat{\mu}_{kq}^{\ell}]^{W}$. This figure suggests that the different layers predominantly exhibit an assortative structure, where nodes tend to interact more with individuals within the same community than with those from different communities. However, we notice some variations for certain layers. For instance, $L_{2}$ (help finding a job) has few non-zero diagonal values, suggesting that this type of support sometimes requires seeking out individuals in different communities. Moreover, $L_{7}$, corresponding to the geographical distance between nodes, has several off-diagonal entries, particularly for communities $K_4$, $K_5$, and $K_6$, suggesting a weakened effect for physical proximity for those communities. Taken together, these findings suggest that \model\ utilizes all the input information to infer partitions that effectively integrate all of them in a meaningful manner. In addition, the inferred affinity matrices illustrate how different layers can exhibit different community structures, a diversity that can be captured by our model.

\end{document}